\documentclass[]{rsos_Ant}

\usepackage{natbib}
\usepackage{graphicx}

\begin{document}

\title{Dust evolution, a global view: \\ IV. Tying up a few loose ends}

\author{
A. P. Jones {\small(Anthony.Jones@ias.u-psud.fr)}}  

\address{Institut d'Astrophysique Spatiale, UMR8617, CNRS, Universit\'e Paris-Saclay, B\^at. 121, 91405 Orsay cedex, France}




\begin{abstract}
This work is an attempt to tie together some of the, seemingly disparate, loose ends from the preceding papers in this series into the global view of dust evolution in the interstellar medium. 
In particular, the nature of the hydro-carbonaceous, infrared absorption and emission band carriers is discussed within the context of a simple poly-hexa-cyclic molecular view. 
The sequence of interstellar dust mantle accretion and grain-grain coagulation into aggregates, and their consequences for the interstellar dust size distribution, extinction and dust thermal emission, is explored. The effects of accretion and coagulation on the interstellar gas-to-dust ratio in the Milky Way and in distant galaxies are estimated. 
The importance of interstellar grain-surface epoxide species is re-visited and the likely and consequent connection between dense interstellar cloud dust and cometary dust explored. 
A possible connection between and the origin of the suite of interstellar OCN-containing molecules, and their interesting chemistry, is discussed. 
Finally, pathways to amino acid formation and their likelihood as a universal basis for life everywhere in the universe is speculated upon. 
\end{abstract}









\begin{center}
{\rhfont  \Huge Dust evolution, a global view: \\ IV. Tying up a few loose ends} 

\vspace{0.8cm}
{\rhfont  \huge A. P. Jones} \vspace{0.2cm} \\ 
{\rhfont (Anthony.Jones@ias.u-psud.fr)} \vspace{0.3cm} \\ 
\noindent {\rhfont  Institut d'Astrophysique Spatiale, CNRS, Universit\'e Paris-Sud, \\ Universit\'e Paris-Saclay, B\^at. 121, 91405 Orsay cedex, France}

\end{center}

\vspace*{1.0cm}

\begin{quote}
{{\bf Abstract:} This work is an attempt to tie together some of the, seemingly disparate, loose ends from the preceding papers in this series into the global view of dust evolution in the interstellar medium. 
In particular, the nature of the hydro-carbonaceous, infrared absorption and emission band carriers is discussed within the context of a simple poly-hexa-cyclic molecular view. 
The sequence of interstellar dust mantle accretion and grain-grain coagulation into aggregates, and their consequences for the interstellar dust size distribution, extinction and dust thermal emission, is explored. The effects of accretion and coagulation on the interstellar gas-to-dust ratio in the Milky Way and in distant galaxies are estimated. 
The importance of interstellar grain-surface epoxide species is re-visited and the likely and consequent connection between dense interstellar cloud dust and cometary dust explored. 
A possible connection between and the origin of the suite of interstellar OCN-containing molecules, and their interesting chemistry, is discussed. 
Finally, pathways to amino acid formation and their likelihood as a universal basis for life everywhere in the universe is speculated upon.} 
\end{quote}


\vspace*{1.5cm}
\noindent {\rhfont \Large 1. Introduction}\\[-1.1cm]
\section{Introduction}
\label{sect_intro}

As per the widely-read works of the galactic hitchhiking author Douglas Adams, this work represents the fourth and, in this case, most definitely the last instalment in the trilogy "Dust evolution, a global view" (Papers I to III). 
So, and to paraphrase Douglas Adams, if you thought that it was a mind-boggingly long way from Paper I to Paper III, then pack your towel, abandon all preconceptions and prepare for a trip to the infinitely-improbable hyperspatial world of interstellar dust chemistry and physics.   

Given the current depth and multi-faceted complexity of our knowledge, that part that we call science is and should remain, of necessity,  borderless, multi-national, apolitical and multi-disciplinary. Over the last century or so the diverse sub-fields of science seem to have become so specialised, and in some cases rather disconnected, that the big picture, a holistic or global view, very often tends gets lost and all too frequently we fail to recognise that the wheel had already been invented long ago. Hence, it is on occasions, interesting to probe less deeply but to explore more widely the shallower waters. 
Dust in the interstellar medium (ISM) may seem a somewhat narrow field of interest but, as will hopefully become clear,  the nature of cosmic dust does have far-reaching consequences. 
With the shallow but wide approach in mind, this paper attempts to find a common denominator answer to the following, seemingly disparate, questions: 
\begin{itemize}
\item What are the carbonaceous, infrared absorption and emission band carriers? 
\item What is the sequence of interstellar dust mantle accretion and coagulation? 
\item Are epoxide species really of any relevance to interstellar chemistry? 
\item What is the connection between interstellar and cometary dust? 
\item Why are OCN\footnote{Molecules, radicals or ions with at least one O, C and N atom that are adjacent to one another, {\it e.g.}, in O--C--N, O=C--N, O=C=N and O--C$\equiv$N bonding configurations.} molecules so common in the ISM? 
\item Why are amino acids the basis of life?  
\end{itemize}
After some initial thoughts, each of the following sections tackles one of these questions before an attempt is made to find a common thread to weave them all together into a coherent picture of cosmic dust, its nature and its evolution. 

The paper is organised as follows: 
Section~\ref{sect_thoughts} presents some preliminary ideas to set the scene for what follows, 
Section~\ref{sect_perceptions} considers some philosophical aspects of interstellar dust modelling, 
Section~\ref{sect_bandcarriers} re-evaluates the nature of the carriers of the interstellar absorption and emission bands in the infrared wavelength region, 
Section~\ref{sect_mantles} discusses carbonaceous mantle formation, dust coagulation and the gas-to-dust ratio in the ISM,  
Section~\ref{sect_ISM_comets} investigates the link between interstellar and cometary dust and chemistry, 
Section~\ref{sect_OCN} points to the fundamental and OCN functional structures to be found within interstellar molecules and how they may relate to dust surface chemistry, 
Section~\ref{sect_epoxides} evaluates the likely roles and importance of epoxide(oxirane)-type functional groups for ISM chemistry,
Section~\ref{sect_amino_acids} makes the link between interstellar chemistry and biochemistry, 
Section~\ref{sect_loose_ends} tries to tie things together into a coherent, global picture of dust and evolution, 
Section~\ref{sect_implications} discusses the broad implications of these diverse threads and, finally,  
Section~\ref{sect_conclusions} concludes this work.

\vspace*{1.5cm}
\noindent {\rhfont \Large 2. Initial thoughts}\\[-1.1cm]
\section{Initial thoughts}
\label{sect_thoughts}

The ideas presented here arise from a re-consideration of the nature of interstellar dust within the framework of the preceding work \citep{2012A&A...540A...1J,2012A&A...540A...2J,2012A&A...542A..98J,2013A&A...558A..62J,2014A&A...565L...9K,2016A&A...588A..43J,2016A&A...588A..44Y,2016RSOS....360221J,2016RSOS....360223J,2016RSOS....360224J} that led to and resulted from the application of the {\em The Heterogeneous dust Evolution Model for Interstellar Solids} \cite[THEMIS,][]{2013A&A...558A..62J,2017A&A...602A..46J}. The previous papers in this series \citep{2016RSOS....360221J,2016RSOS....360223J,2016RSOS....360224J} left a few threads hanging and 
so this work is an attempt to plug those gaps and to add some essential filler. 
Unlike other widely-used dust  models the THEMIS approach\footnote{http://www.ias.u-psud.fr/themis/ THEMIS is based on the chemical and physical properties of interstellar dust analogue materials directly-measured in the laboratory or built ground-up by interpretation and extrapolation of laboratory measurements.}  
encompasses the view that that the dust properties respond to and, in turn, influence their surroundings. For example, it is no longer sufficient to interpret the observed variations in the dust properties in the ISM as simply a change in the size distribution, instead the observed variations must be seen in terms of a coupled evolution of the dust chemical composition, size and shape distribution and structure ({\it e.g.}, multi-layer mantles, coagulated/aggregates, {\it etc.}). 
This is the basic premiss of the THEMIS dust modelling approach and for all that follows here. 



\newpage
\noindent {\rhfont \Large 3. Misguided perceptions and preconcpetions}\\[-1.1cm]
\section{Misguided perceptions and preconcpetions}
\label{sect_perceptions}

Seemingly and all too often our perceptions\footnote{Perception ({\it noun})  - 1. the ability to see, hear, or become aware of something through the senses. 2. the way in which something is regarded, understood, or interpreted.} and preconceptions\footnote{Preconception ({\it noun}) - a preconceived idea or prejudice.} interfere in our efforts to objectively construct a viable view of the Universe. This usually arises because we are too vested in the currently-accepted view of things and therefore fail to adequately re-asses or re-appraise those views in the light of new evidence. Rather than re-write the book it is usually much simpler to tweak the current view to fit the observations. 
However, the continual tweaking of a world view -- an empirical approach -- often ends up losing sight of the original goals of such a view.  Under such circumstances it is  better to abandon the old views and to start over with a fresh, more enlightened and satisfying view. Just because a current model is successfully able to fit a given set of observations doesn't necessarily mean that that model is close to reality. The real test of a model comes when it is confronted by new observations. A model that can match the new observations without tweaking or tuning is indeed on the road to success but still needs to be constantly tested against new data as they become available for comparison. 
Models that fail to meet the tests laid before them by new data should clearly be abandoned. 
In the following a few key examples are given of how our world view of cosmic dust has failed to fully appreciate and to incorporate some physical realities. 

It appears that the burgeoning results from the field of nano-particle physics have generally not been sufficiently-considered within the context of interstellar dust particles, which is particularly ironic because practically all of these grains are, literally, nano-particles. 
The physical and chemical properties of nano-particles (here defined as particles with radii $\lesssim 50$\,nm) are usually and simply derived by extrapolation of bulk material properties (particles with radii $\gtrsim 100$\,nm). This classical approach singularly fails to account for particle size limitations and important effects at the surfaces of nano-structures, which often result in enhanced reaction rates and spectacular catalytic reactions not possible with larger particles. 
An example of an important particle size effect concerns our myopic view of carbonaceous interstellar dust analogue materials which, even if of a completely aromatic nature, cannot be as black as soot, as is implicitly assumed in the extrapolation of bulk material properties to nanometre scales. At large size-scales a grain can indeed be black, for example comets, includng 67P/Churyumov-Gerasimenko, have very low albedos ($0.02-0.06$) due to the photolysis of interstellar matter into aromatic-rich and hard bulk materials. However, at sub-micron and especially at nanometre sizes this no longer holds true because the highly absorbing aromatic domains are limited in size by the particle dimensions, leading to a systematic shift in the absorption away from visible light-absorbing dark materials to particles that are practically transparent in the visible but absorb strongly in the ultraviolet. 

The interstellar polycyclic aromatic hydrocarbon (PAH) hypothesis has, since its proposal many decades ago, received much observational, experimental and theoretical attention, yielding innumerable interesting results in the process. However, the existence of planar, completely aromatic molecules in the ISM has not yet been proven and much of the recent work from the "PAH community" appears to be somewhat eroding the idealistic PAH supposition. Indeed some work seems to be indicating that the IR emission band carrier species are probably rather messed-up and far from the ideal, classical PAHs \citep[{\it e.g.},][]{2010A&A...510A..36M,2015PhRvA..92e0702G,2015arXiv150101716O,2015A&A...577A..16P,2016ApJ...832...24W}.   
In fact, and to date, not a single PAH has been detected in the ISM. However, pure aromatics, such as fullerenes \citep[PAs,][]{2010Sci...329.1180C} and cyclic aromatics aromatics, such as benzene and cyanobenzene \citep[AHs,][]{2001ApJ...546L.123C,2018Sci...359..202M}, have indeed been observed. Thus, it would appear that it is time to objectively revisit the PAH hypothesis and make a reappraisal of its usefulness. Additionally, it now looks like the detection of fullerenes in space is leading to a lurch in a fullerenes-explain-all direction, which is ultimately probably going to prove about as useful as the major investments that have been made in the PAH hypothesis. In conclusion, while PAHs and fullerenes are interesting species and likely to exist in space, their abundances are dwarfed by the bulk of the carbonaceous matter that is not in either of these rather particular forms. 

Another key issue is the formation route to molecular hydrogen in the ISM. This is classically taken to occur via the addition reactions of physisorbed H atoms on grain surfaces \citep[{\it e.g.},][]{1970JChPh..53...79H,1971ApJ...163..155H,2014A&A...569A.100B,2017MolAs...9....1W}. However, in photo-dissociation regions (PDRs) this mechanism fails because H atom residence times are too short. Thus, an alternative mechanism in regions where the dust is hotter is needed, and has indeed been known of for almost 20 years \citep{1984JAP....55..764S,1989JAP....66.3248A,1996MCP...46...198M}. This work shows that irradiation, thermal or photo-processing of hydrogenated amorphous carbonaceous materials liberates H$_2$ under the effects of radiolysis, thermal annealing and/or photolysis and could therefore also be of critical importance in the ISM \citep{2014A&A...569A.119A,2015A&A...581A..92J}. 

Other than surface chemistry leading to H$_2$ formation in the ISM, little thought has yet been given to the active chemistry that can take place on amorphous hydrocarbon grain surfaces. A topic that was dealt with in some detail on the previous papers in this series \citep{2016RSOS....360221J,2016RSOS....360224J} and is again looked at later in this paper. Thus, routes to simple radicals and simple and complex molecule formation on the surfaces of abundant nano-particles in the ISM may be missing significant reaction routes that could, depending on the local physical conditions, end up being more efficient than formation by gas phase chemistry.  

In conclusion, in our re-thinking of interstellar dust in all its manifestations we need to use our imaginations to the full and to adopt a significant degree of lateral thinking.

\vspace*{1.5cm}
\noindent {\rhfont \Large 4. The carbonaceous nature of the infrared band carriers}\\[-1.1cm]
\section{The carbonaceous nature of the infrared band carriers }  
\label{sect_bandcarriers}

The most physically-realistic interstellar carbonaceous grain analogue materials would appear to be the extensive family of (hydrogenated) amorphous carbons, a-C(:H). These semiconducting materials span a wide compositional range, from a-C, aromatic-rich materials with a low H atom fraction ($\lesssim 15$\%) and narrow optical gap ($E_{\rm g} < 1$\,eV), to a-C:H, aliphatic-rich materials with a high H atom fraction ($\sim15-60$\%) and wide optical gap ($E_{\rm g} \geq 1$\,eV). 
Hydrogenated amorphous carbons, a-C(:H), are macroscopically-structured, contiguous network, solid-state materials comprised of only carbon and hydrogen atoms whose properties have been well-studied within both the physics and astrophysics communities \citep[{\it e.g.},][]{1979PhRvL..42.1151P,1980JNS...42...87D,1983JNCS...57..355T,1986AdPhy..35..317R,1987PhRvB..35.2946R,1988JVST....6.1778A,1988Sci...241..913A,1988PMagL..57..143R,1990JAP....67.1007T,1991PSSC..21..199R,1995ApJS..100..149M,1996ApJ...464L.191M,2000PhRvB..6114095F,2001PSSAR.186.1521R,2002MatSciEng..37..129R,2003ApJ...587..727M,2004PhilTransRSocLondA..362.2477F,2007DiamondaRM...16.1813K,2007Carbon.45.1542L,2008ApJ...682L.101M,2011A&A...528A..56G,2012A&A...540A...1J,2012A&A...540A...2J,2012A&A...542A..98J}. 

In paper II of this series \citep{2016RSOS....360223J} it was proposed that the intrinsic structures of asphaltenes, the natural, aromatic-rich moities found within crude petroleum and coal, could be analogues of the sub-structures within aromatic-rich, X-atom doped, hydrogenated amorphous carbon,\footnote{} a-C(:H[:X]). 
In particular, it was suggested that such interstellar nano-particle sub-structures could be responsible for the diffuse interstellar bands (DIBs), where the doping X atoms are principally N, O and/or S but may also include Si, Ge, Sn, P, As and B atoms \citep{2013A&A...555A..39J,2014P&SS..100...26J,2016RSOS....360223J}.  
Asphaltenes, of whatever origin, show a wide range of complex structures, often with peripheral methyl and larger alkyl  groups. Their aromatic domains tend to be relatively small, with predominantly $4-10$ sixfold rings but also include a significant fraction of fivefold rings and rare sevenfold rings.  
Indeed, such species had already been suggested as an analogue model for interstellar carbonaceous grains. For example, it was proposed that the emission spectra of the dust observed in proto-planetary nebul\ae\ (PPNe) could be explained by carriers that resemble the complex chemical structures in the highly aromatic fractions (asphaltenes) found within naturally-occurring coals and crude oil \citep{1989A&A...217..204P,1991A&A...247..215P,2004OLEB...34...13C,2013AcSpA.111...68C}. The inherent structures of these materials are not purely aromatic but consist of an intimate mix of aromatic domains and cyclic aliphatics, which is consistent with observations indicating that the IR emission bands originate in particles containing a mix of aromatic, olefinic and aliphatic species \citep{2012A&A...540A...1J,2012A&A...540A...2J,2012A&A...542A..98J,2013A&A...558A..62J}. It appears that the best match to the PPNe emission bands in the $2.5-200\,\mu$m wavelength region is for the asphaltenes derived from the heavy oil fraction "{\it which are composed by an aromatic core containing three to four condensed aromatic rings surrounded by cycloaliphatic (naphtenic) and aliphatic alkyl chains.}" \citep{2013MNRAS.429.3025C}. 

In paper II of this series \citep{2016RSOS....360223J} a supposedly new carbon belt-like molecule, (C$_4$H$_2$)$_n$, was proposed (and named couronene) but, as is probably too often the case in today's literature-overloaded world, such molecules, along with many other more interesting and complex "Molecular Belts and Loops", had been described ten years earlier \citep{Tahara:2006er} and named n-cyclacenes by the authors.\footnote{The author humbly apologises to the authors Tahara and Tobe \citep{Tahara:2006er} for having failed to notice their prior claim on this molecular species and wishes to thank Olivier Bern\'e for having pointed out this omission.} 

\begin{figure}[!h]
\centering\includegraphics[width=5.0in]{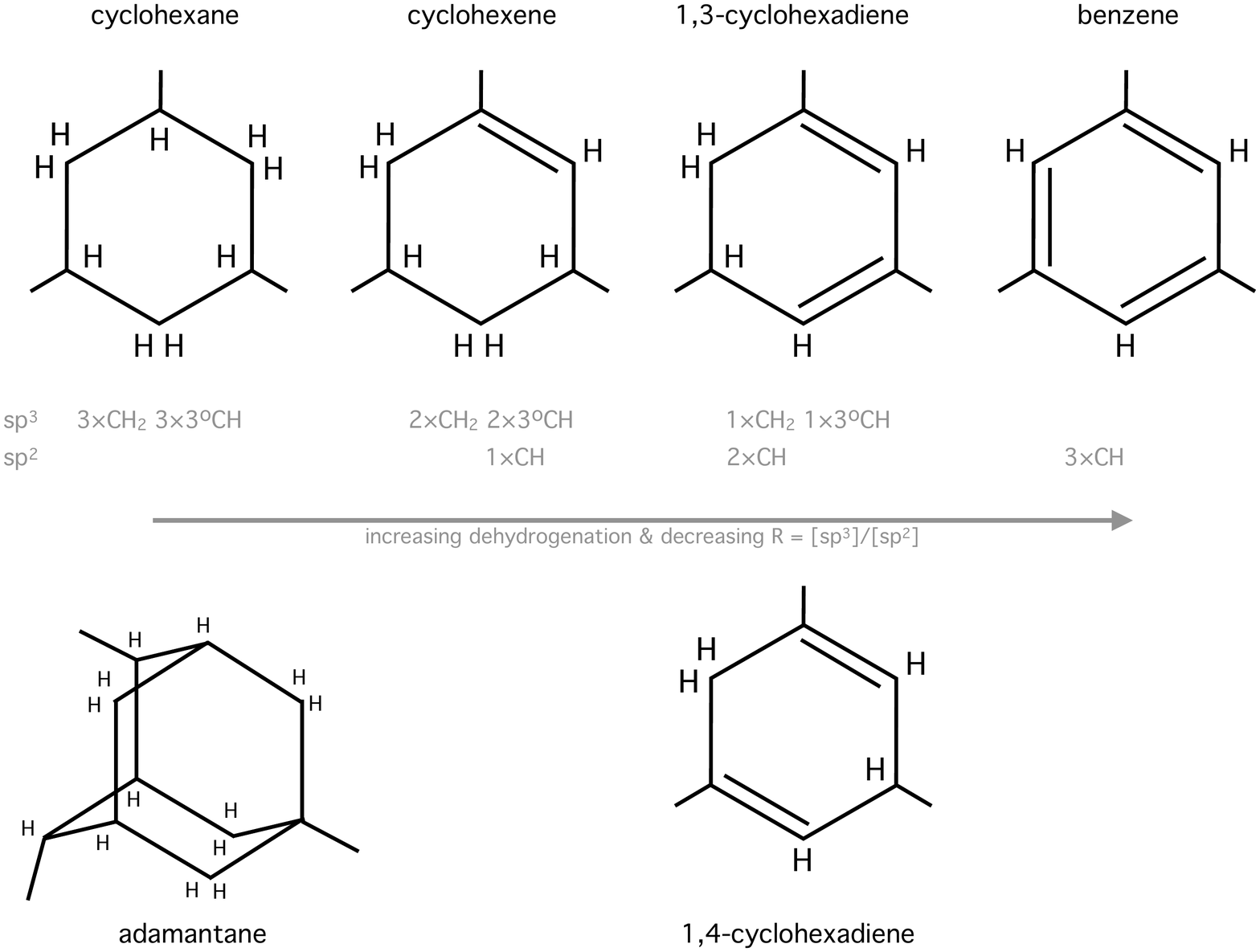}
\caption{An illustrative dehydrogenation evolution scenario for six-fold cyclic structures within a-C(:H) materials, starting with a completely aliphatic cyclohexane-like ring and evolving through the increasingly olefinic cyclohexene-like  and cyclohexadiene-like and ending with the fully aromatic benzene-like cycle. The adamantane structure is shown for comparison. In each case it is assumed that the six-fold ring is part of the same contiguous structure to which it is attached by three trigonally-arranged C$-$C bonds (shown as lines external to the six-fold rings), except in the case of adamantane where the external bonds are not exactly trigonally-arranged. The single H atoms inside (outside) the six-fold rings indicate tertiary, $3^\circ$, aliphatic sp$^3$ (olefinic or aromatic sp$^2$) CH groups. The paired H atoms outside the six-fold rings indicate sp$^3$ CH$_2$ groups.}
\label{fig_cyclics}
\end{figure}

\begin{figure}[!h]
\centering\includegraphics[width=4.4in]{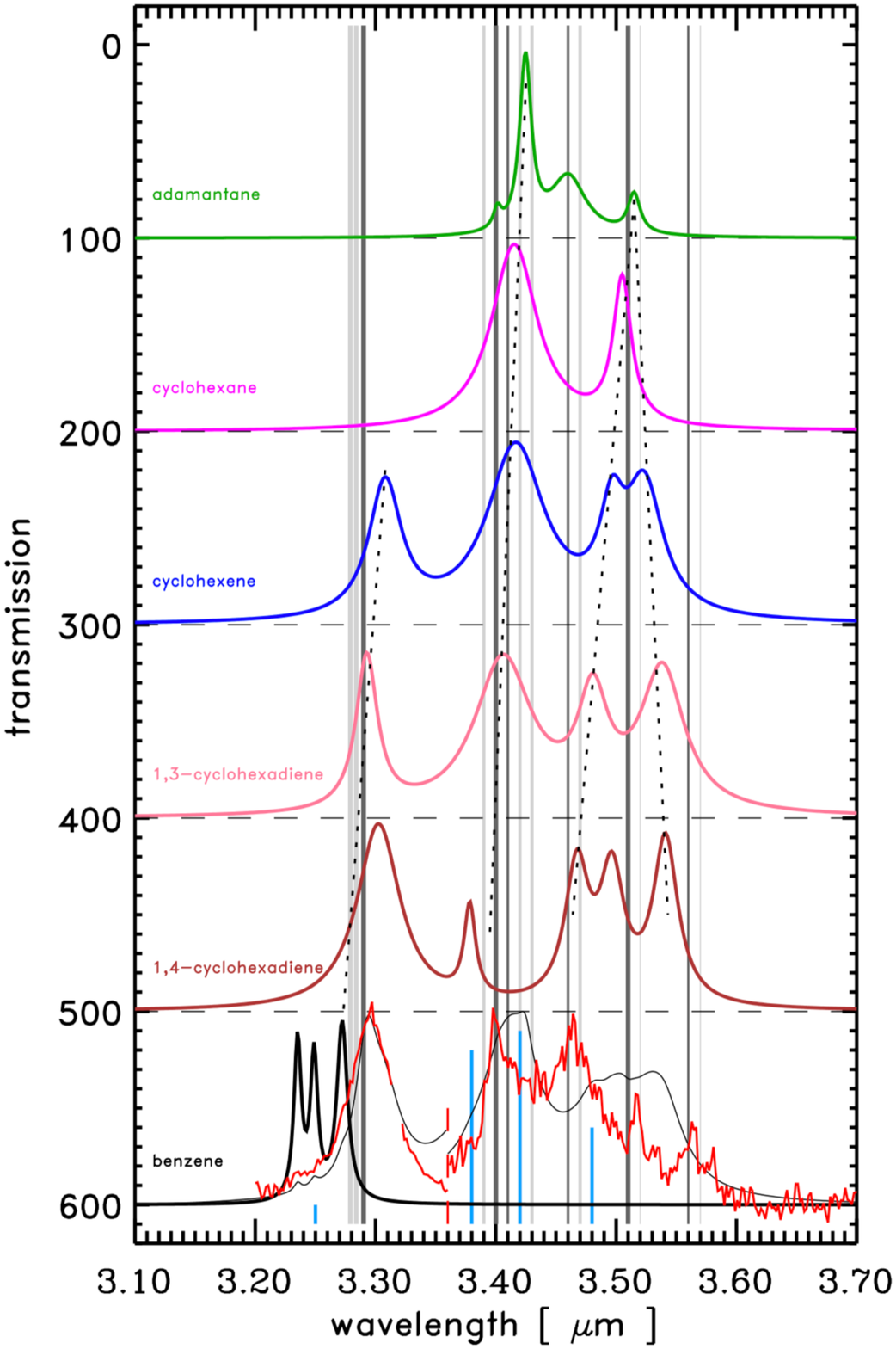}
\caption{A schematic comparison of the $\lambda = 3-4\,\mu$m spectra of the polycyclic aliphatic molecule adamantane, the aliphatic cyclohexane, the mixed olefinic/aliphatics cyclohexene, 1,3-cyclohexadiene and 1,4-cyclohexadiene and the fully aromatic molecule benzene. 
The thin grey line overlying the benzene spectrum is a linear combination of the six spectra (in the same order as above) in the ratio 
0.14 : 0.27 : 0.41 : 0.03 : 0.03 : 0.14 
and schematically illustrates what a mixed component spectrum could look like. 
The red line shows the IR emission band spectrum of the dust in the disc around HD100546 \citep{Bouteraon_etal_2018}.  
The latter spectra are, for clarity, scaled separately in the wavelength regions less than and greater than $3.36\,\mu$m. 
The vertical dark grey lines indicate the central positions of the observed IR emission bands and the thin grey lines mark the positions of bands in laboratory-measured samples, as described by \cite{Bouteraon_etal_2018} (and references therein). The short light blue lines schematically-indicate the positions and intensities of the IR bands observed in absorption towards the Galactic Centre \citep{2002ApJS..138...75P}. The slanting short-dashed lines indicate the band shifts discussed in the text.}
\label{fig_cyclics_spectra}
\end{figure}

An understanding of the mechanistic details of the photo-processing or photolysis of amorphous hydrocarbon materials is a key to advancing our studies of how these materials form and evolve under ISM conditions \citep[{\it e.g.},][]{2008A&A...490..665P,2012A&A...540A...1J,2012A&A...540A...2J,2012A&A...542A..98J,2014A&A...569A.119A,2015A&A...577A..16P,2015A&A...584A.123A}. In this case it perhaps pays to go back to the basics and consider the building blocks of a-C:H solids. It is clear that (poly-)cyclic species are an important component in their structures \citep[{\it e.g.},][]{1986AdPhy..35..317R,1988PMagL..57..143R,2012A&A...540A...1J,2012A&A...540A...2J,2012A&A...542A..98J,2013MNRAS.429.3025C}. Thus, if we consider isolated six-fold ring systems, starting with  saturated adamantane- or cyclohexane-type structures, we can study how these simple systems are likely to evolve as a result of UV photon-induced dehydrogenation. A likely sequence for this photolytic processing is shown in Fig~\ref{fig_cyclics}, where step-wise dehydrogenation transforms a single six-fold aliphatic cycle into an aromatic benzene-type structure with the removal of a pair of H atoms in each step. Experimental evidence points to the removal of pairs of atoms as molecular hydrogen under the effects of photolysis, ion irradiation and thermal annealing \citep{1984JAP....55..764S,1989JAP....66.3248A,1996MCP...46...198M,2014A&A...569A.119A}. 
Given that IR spectra are well-determined, diagnostic chemical fingerprints for the CH$_n$ ($n = 1-3$) vibrational modes in hydrocarbons we can compare their $\lambda = 3-4\,\mu$m spectra in order to elucidate the indicative signatures that can help to resolve the CH bonding structures likely to found within a-C(:H) nano-grains in the ISM. 
In order to do this we consider the attribution of the IR CH stretching modes in the $3-4\,\mu$m region using a recent graphically-illustrated comparative compilation \cite[][see Fig. 1 in this work and the references cited therein]{Bouteraon_etal_2018}. 
The spectra of the considered hexa-cyclic molecular species are shown in Fig.~\ref{fig_cyclics_spectra}, along with that of the poly-hexa-cyclic aliphatic molecule adamantane (C$_{10}$H$_{16}$). The data used in this re-analysis are summarised in Table~\ref{table_cyclos} (IR band assignments) and Table~\ref{table_cyclo_struc} (IR band analysis) and discussed in detail in the following.  
The bands have been grouped by their central wavelengths, to within at most $\pm 0.02\,\mu$m in order to give some leeway and allow for the fact that the exact band centres for a particular CH$_n$ ($n = 1-2$) configuration will depend upon the local environment in the molecule (or a-C(:H) sub-structure). 
Table~\ref{table_cyclos} and Fig.~\ref{fig_cyclics_spectra} indicate that the spectra of these six molecules broadly exhibit six groups of bands (with their general intensity and shape characteristics), {\it i.e.},
\begin{enumerate}
\item $3.25\pm 0.02\,\mu$m   \hspace*{0.3cm} (strong, narrow),
\item $3.30\pm 0.01\,\mu$m   \hspace*{0.3cm} (strong, broad), 
\item $3.40\,\mu$m                 \hspace*{1.25cm} (weak, narrow),  
\item $3.42\pm 0.01\,\mu$m   \hspace*{0.3cm} (strong, broad; strong, narrow for adamantane),  
\item $3.46\,\mu$m                 \hspace*{1.25cm} (moderate, narrow), 
\item $3.50\pm 0.02\,\mu$m   \hspace*{0.3cm} (strong, moderately broad; weak, narrow for adamantane) and  
\item $3.53\pm 0.01\,\mu$m   \hspace*{0.3cm} (strong, moderately broad).  
\end{enumerate}
It is notable that the dispersions in the band centres are, at most, $\pm 0.02$ and in most cases only $\pm 0.01$. This gives some confidence that the so-grouped bands in this suite of molecules can indeed be assigned to particular CH$_n$ group configurations. 
It is also notable that the positions of these bands, if not their intensities, bear a remarkable resemblance to the positions of the observed IR emission bands (also shown in Table~\ref{table_cyclos}). 
Certain of these IR stretching bands are clearly related to well-established and sound attributions \citep[{\it e.g.}, see][]{Bouteraon_etal_2018} and may be assigned to   
\begin{itemize}
\item $3.29\,\mu$m aromatic CH \hspace*{1.2cm} ($\sim 3.25\,\mu$m, benzene) 
\item $3.42\,\mu$m asymmetric CH$_2$  \hspace*{0.65cm} ($3.42\pm 0.01\,\mu$m) 
\item $3.47\,\mu$m tertiary aliphatic CH \hspace*{0.12cm} ($3.46\,\mu$m, adamantane)  
\item $3.50\,\mu$m symmetric CH$_2$  \hspace*{0.8cm} ($3.50\pm 0.02\,\mu$m).     
\end{itemize}
Among the six molecules studied here the three narrow $\simeq 3.25\,\mu$m aromatic CH bands only appear in benzene. A broader band at these same wavelengths appears to be a characteristic of the spectra of co-added fully aromatic molecules \citep{2013ApJS..205....8S}, which does not appear to be consistent with the hypothesised existence of fully aromatic molecules in space (the so-called "interstellar PAHs"). 
Other than benzene, the most conjugated of the six molecules, 1,3-cyclohexadiene, with a $-$C$=$C$-$C$=$C$-$ conjugation within the ring, exhibits an olefinic CH stretching vibration at $\simeq 3.29\,\mu$m, which is consistent with the position and the width of the $3.29\,\mu$m band observed in a stellar disc \citep{Bouteraon_etal_2018} (see Fig.~\ref{fig_cyclics_spectra}), and with the interstellar $3.29\,\mu$m emission band.  The fit in Fig.~\ref{fig_cyclics_spectra} is indeed much closer to the observed emission band than any co-added fully aromatic mix \citep{2013ApJS..205....8S}.   
Additionally, both cyclohexene and 1,4-cyclohexadiene also show bands at similar wavelengths, {\it i.e.}, $3.31$ and $3.30\,\mu$m, respectively. 
A satisfactory fitting of this key material-characteristic IR band within the context of "interstellar PAH" modelling is unfortunately often not given sufficient weight. 
Thus, and despite the very widely-held view, it appears that highly aromatic molecules may not be the best interpretation of the observed $3.29\,\mu$m IR emission band, which is in any event likely composed of two sub-bands \citep{2003MNRAS.346L...1S} of possible sp$^2$ olefinic and/or aromatic origin \citep{2017ApJ...845..123S,Bouteraon_etal_2018}.

\vspace*{1.5cm}
\noindent {\rhfont \Large 4.1. Infrared band analysis and interpretation}\\[-1.1cm]
\subsection{Infrared band analysis and interpretation} 
\label{sect_spect_analysis}

\begin{table}[!h]
\caption{The IR band assignments for the  C$-$H stretching modes in the hexa-cyclic molecules shown in Fig~\ref{fig_cyclics}.
Note that adamantane is a polycyclic aliphatic molecule with four interlocked hexa-cycles and that benzene is the only fully-aromatic molecule. See the text for a justification of the band assignments, where aromatic (ar.) and olefinic (ol.) sp$^2$ origins are indicated. Where there is any ambiguity the H atoms assumed to be responsible for the given IR bands in each of the CH$_n$ configurations are shown in boldface ({\bf H}). All the tabulated bands are comparatively strong, except for the three adamantane bands shown it {\it italic}. The positions of the observed interstellar emission bands are also indicated, with the most prominent bands shown in {\bf boldface}; the larger uncertainties for the $3.46$ and $3.56\,\mu$m bands reflect the difficulty in determining the band centres for these weaker and broader bands. 
}
\label{table_cyclos}
\begin{center}
\begin{tabular}{lccccccc}
\hline
                 &        &      &                             &               &               &        &       \\[-0.3cm]
&  \multicolumn{7}{c}{band centre wavelength [ $\mu$m ]} \\[0.2cm] 
 \hspace{1.2cm} origin $\rightarrow$        & sp$^2$  & sp$^2$ &  sp$^3$   & sp$^3$  &  sp$^3\pm$\,sp$^2$  &    sp$^3$ &  sp$^3$+\,sp$^2$  \\[0.05cm]
   & ar.                     &  ol.                    & $2^\circ$ CH$_2$ & C{\bf H}$_2$C{\bf H}$_2$ & $\geqslant$CH &  & \\[0.05cm]
   & $\geqslant$CH & $\geqslant$CH & $3^\circ$ CH    & C{\bf H}$_2$C{\bf H}$<$ & $3^\circ$ CH        & $2^\circ$CH$_2$ & C{\bf H}$_2$CH$=$ \\[0.05cm]
    molecule                                               &    (1)      &     (2)    &    (3)         &    (4)                        &     (5)           &       (6)    &    (7)     \\
                                                                 &               &             &                  &              &                &                    &        \\[-0.3cm]
\hline
                  &      &      &       &     &   &    \\[-0.3cm]  
 adamantane              &            &          &  {\it 3.401}   &  3.425   &   {\it 3.460}   & {\it 3.515}   &            \\
 cyclohexane              &           &           &              &  3.415   &             & 3.505   &           \\
 cyclohexene              &           &  3.308  &            &  3.416   &             & 3.497   &  3.524  \\
 1,3-cyclohexadiene   &            &  3.293 &            &  3.406   &             & 3.481   &   3.539   \\
 1,4-cyclohexadiene   &            &  3.303 &            &  (3.378) &   \multicolumn{2}{c}{3.467}    &  3.541    \\
 \hspace{0.5cm}   "     &            &            &            &            &              & 3.497    &              \\
 benzene                    &    3.235  &         &            &            &             &             &           \\
 \hspace{0.5cm}   "     &    3.249  &         &            &            &             &             &           \\
 \hspace{0.5cm}   "     &    3.272  &         &            &            &             &             &           \\
        &       &      &       &        &      &     &    \\[-0.3cm]
\hline
        &       &      &       &        &      &     &    \\[-0.3cm]
 average                      &    3.25           &  3.30           &  3.40           &   3.42          &  3.46          &  3.50           & 3.53             \\
 $\approx$ dispersion  &   $\pm0.02$  &  $\pm0.01$ &                    & $\pm0.01$  &                  &  $\pm0.02$ &  $\pm0.01$  \\[-0.3cm]

 & & & &  & & & \\\cline{4-5}
        &       &      &       &        &      &     &    \\[-0.3cm]
\hline
        &       &      &       &        &      &     &    \\[-0.3cm]
 IS emission bands                        &   &  {\bf 3.29} & \multicolumn{2}{c}{\bf 3.40}          & 3.46        &  {\bf 3.51}    &  $\sim 3.56$  \\
 $\approx$ uncertainty    &   &  $\pm0.01$  & \multicolumn{2}{c}{$\pm0.01$} & $\pm0.02$  & $\pm0.01$ &  $\pm0.03$ \\[0.05cm]
\hline
        &       &      &       &        &      &     &    \\[-0.25cm]
\end{tabular}  
\begin{list}{}{}   
Band assignments:
\item[] 1). Aromatic $\geqslant$CH modes in the highly symmetric molecule benzene. 
\item[] 2). Olefinic $=$C$<^{\rm H}$ mode. \ {\it N.B.}, not an olefinic $=$C$<^{\rm H}_{\rm H}$ group.  
\item[] 3). A weak aliphatic secondary ($2^\circ$) $>$CH$_2$ or tertiary ($3^\circ$) $>$CH$-$ mode. 
\item[] 4). Adjacent $2^\circ$ and $3^\circ$ aliphatic CH bonds: $-$CH$_2-$CH$_2-$ and $-$CH$_2-$CH$<$.
\item[] 5). $3^\circ$ sp$^3$ $>$CH$-$ or an olefinic $=$C$<^{\rm H}$ next to an sp$^3$ CH$_n$ ($n = 1,2$), {\it e.g.} $=$C$<^{\rm H}_{\rm CH_2-}$.
\item[] 6). Isolated aliphatic $>$CH$_2$. 
\item[] 7). Olefinic $=$C$<^{\rm H}$ next to an aliphatic CH$_n$, {\it i.e.} $=$C$<^{\rm H}_{\rm CH_2-}$ \ or \ $=$C$<^{\rm H}_{\rm CH<}$.
\end{list}    
\end{center} 
\end{table}

\begin{table}[!h]
\caption{An analysis of the CH bonding characteristics of isolated hexa-cyclic molecules and the analogous hexa-cyclic structures within a contiguous a-C(:H) network where three of the CH bonds have been replaced with trigonally arranged network-linking C$-$C bonds (see Fig~\ref{fig_cyclics}).}
\label{table_cyclo_struc}
\begin{center}
\begin{tabular}{lcccccc}
\hline
                 &        &      &                             &               &               &               \\[-0.25cm]
&  \multicolumn{6}{c}{No. of groups per molecule ( per a-C(:H) sub-structure )} \\[0.2cm] 
                 &  adjacent   &              &   3$^\circ$ &  aliphatic &              &               \\ 
                 &  aliphatic   & alphatic &  aliphatic &  olefinic &  olefinic      &  aromatic \\
 molecule & CH$_2$CH$_{1,2}$   & CH$_2$  &        CH   &    CH$_{1,2}$CH=    & =C$<^{\rm H}$   &    CH        \\
                 &               &          &                   &             &                &               \\[-0.3cm]
\hline
                  &      &      &       &     &   \\[-0.25cm]  
 adamantane              &   12 ( 5 )  &  6 ( 4 ) &  4 ( 5 ) &         &            &              \\
 cyclohexane              &   6 ( 6 )  &  6 ( 3 ) &  -- ( 3 ) &           &            &              \\
 cyclohexene              &   3 ( 3 )  &  4 ( 2 ) &  -- ( 2 ) & 2 ( 1 ) &            &             \\
 1,3-cyclohexadiene   &   2 ( 1 )  &  2 ( 1 ) &  -- ( 1 ) & 2 ( 2 )  &  4 ( 2 ) &              \\
 1,4-cyclohexadiene   &               &  2 ( 1 ) &  -- ( 1 ) & 4 ( 2 ) &  4 ( 2 ) &             \\
 benzene                    &            &               &             &                             &             &  6 ( 3 )  \\[0.05cm]
\hline
        &       &      &       &        &      &     \\[-0.25cm]
 IR band(s) [$\mu$m] &  3.50  & 3.42 & 3.40, 3.46 & 3.30, 3.53 & 3.30 &  $\sim 3.25$  \\[0.05cm]
\hline
        &       &      &       &        &      &     \\[-0.25cm]

\end{tabular}     
\end{center} 
\end{table}

This section considers the interpretation of the origins of the signature IR bands and, where there appears to be some ambiguity, re-considers some of those assignments within the context of the six cyclic molecules considered here. 

The $3.42\,\mu$m band is almost certainly of aliphatic CH$_2$ group origin, as appears to be widely accepted.  
However, the fact that 1,4-cyclohexadiene contains two isolated CH$_2$ groups and does not show a $3.42\,\mu$m band must call this assignment into some doubt. Nevertheless, 1,4-cyclohexadiene does have a weaker band at $3.47\,\mu$m (see Fig.~\ref{fig_cyclics_spectra}).  
Here we propose that the $3.42\,\mu$m band is indeed due to aliphatics but that it is due to a coupled vibrational mode arising from adjacent (at least paired)   CH$_2$ groups.
This same band which is seen in adamantane, where there are no paired CH$_2$ groups, means that this band may also arise from aliphatic CH$_2$ groups adjacent to a tertiary (3$^\circ$) aliphatic CH bonds, sites that are present in adamantane.
It is therefore proposed that the $3.42\,\mu$m band is actually due to $-$CH$_2$$-$CH$_2-$ and $-$CH$_2-$CH$<$ structural groups. 
Thus, the interstellar $3.40-3.42\,\mu$m band would appear to be a signature of cyclic aliphatics or of the aliphatic bridges that link olefinic/aromatic domains within a-C(:H) nano-grains \citep{2012A&A...542A..98J,2012ApJ...761...35M,2015A&A...581A..92J}. 
The $3.42\,\mu$m band in adamantane is much narrower than in the cyclohexenes but it, and its satellite bands at 3.40 and $3.46\,\mu$m merged together, are probably equivalent to the broader band seen in cyclohexene, 1,3-cyclohexadiene and 1,4-cyclohexadiene (see Fig~\ref{fig_cyclics_spectra}).   

The $3.46\,\mu$m band is usually identified with tertiary (3$^\circ$) aliphatic CH bonds \citep{1992ApJ...399..134A}, however, this unique assignment might appear shaky given that a band at $3.47\,\mu$m occurs in 1,4-cyclohexadiene, which contains no 3$^\circ$ CH configuration. 
Nevertheless, and as we discuss below, the $3.47\,\mu$m band in 1,4-cyclohexadiene is probably a 2$^\circ$ aliphatic CH$_2$ transition shifted to shorter wavelengths due to strain in the mixed sp$^3$-sp$^2$ six-fold ring. 

The $3.50 \pm 0.02\,\mu$m band must be associated with solo aliphatic CH$_2$ groups because it is observed in adamantane, cyclohexane, cyclohexene, 1,3-cyclohexadiene and 1,4-cyclohexadiene, all of which contain isolated aliphatic CH$_2$ groups but only three of these five molecules contain adjacent CH$_2$ groups.

The $3.53 \pm 0.02\,\mu$m band appears in all, and only in, the three cyclohexenes and it is here proposed that this band is another coupled mode but in this case due to the vibration of an aliphatic CH$_2$ group next to an olefinic CH bond, {\it i.e.}, it arises in a $-$CH$_2-$CH$=$ structure.    

A close look at the schematic spectra in Fig.~\ref{fig_cyclics_spectra} reveals some interesting trends that depend upon the olefinic bond content and the degree of single-double bond conjugation in mixed aliphatic/olefinic molecules, which is generally associated with increasing ring strain as the olefinic to aliphatic ratio increases. These trends are highlighted by the slanted short-dashed lines in Fig.~\ref{fig_cyclics_spectra}, which link bands assumed to be of the same origin. 
For example, the peak of the $\simeq 3.3\,\mu$m band due to olefinic CH bonds shifts in wavelength from $3.30$ to $3.29\,\mu$m with increasing conjugation. Note that 1,4-cyclohexadiene (peak at $3.30\,\mu$m) and 1,3-cyclohexadiene (peak at $3.29\,\mu$m) each contain two olefinic C$=$C bonds but that they are only conjugated in 1,3-cyclohexadiene, {\it i.e.}, $-$C$=$C$-$C$=$C$-$.
The same band peaks at $3.31\,\mu$m in cyclohexene where there is only a single C$=$C bond, {\it i.e.}, close to the band centre in 1,4-cyclohexadiene.  
In the fully aromatic benzene molecule this same CH stretching band appears at shorter wavelengths still, {\it i.e.}, $\simeq 3.23-3.27\,\mu$m. 
A similar shift in the aliphatic CH$_2$ band at $\sim 3.4\,\mu$m is also seen; in this case the band shifts from $3.43$ to $3.38\,\mu$m, with decreasing aliphatic content. 
Another clear trend that can be discerned is that the $\sim 3.5\,\mu$m band appears to split into two principal components with increasing olefinic content, one band shifts to shorter wavelengths (from $3.51$ to $3.47\,\mu$m), while other shifts to longer wavelengths (from $3.51$ to $3.54\,\mu$m). In this case the band shifts would appear to be roughly symmetric with respect to $3.51\,\mu$m. 

The composite spectrum, dominated by cyclohexane and cyclohexene,  
shown by the thin grey line in the bottom part of Fig.~\ref{fig_cyclics_spectra} is compared with the spectrum of the dust in the disc around HD100546 (red line) \citep{Bouteraon_etal_2018}. 
The band position and shape of the $3.30\,\mu$m band is remarkably well-fit here with a mixture that contains no aromatic species but which is comprised of the sub-bands of cyclohexene and 1,3-cyclohexadiene at $3.31$ and $3.29\,\mu$m, respectively.  
Qualitatively, the $3.4-3.5\,\mu$m wavelength region fit, shown in the lower spectrum in Fig.~\ref{fig_cyclics_spectra}, is reasonable but this molecular mix is unable to explain the $3.46$ and $3.56\,\mu$m bands. This spectral mis-match is re-visited in the following sub-section. 

\begin{figure}[!h]
\centering\includegraphics[width=4.75in]{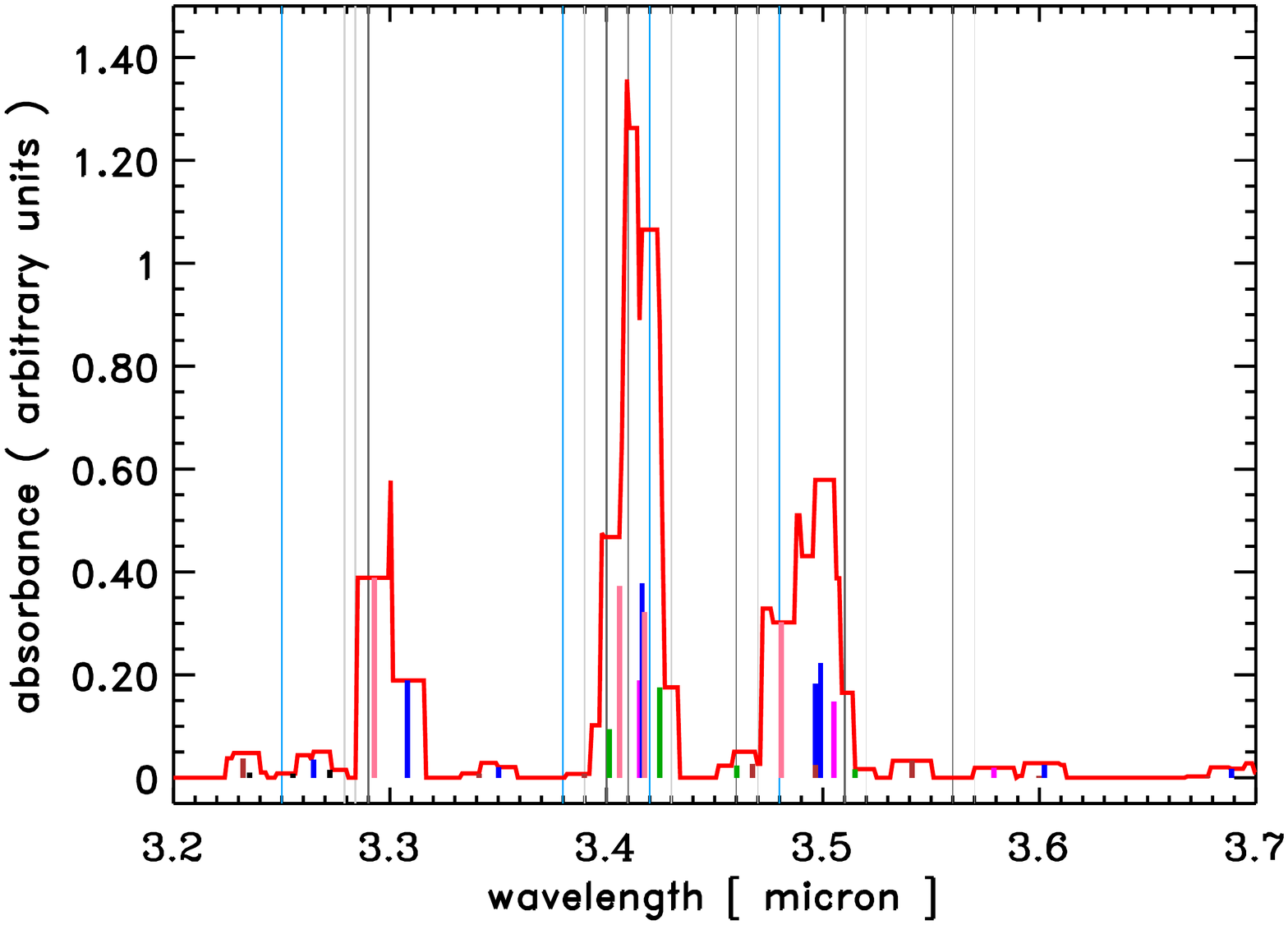} \\[-1.0cm]
\centering\includegraphics[width=4.75in]{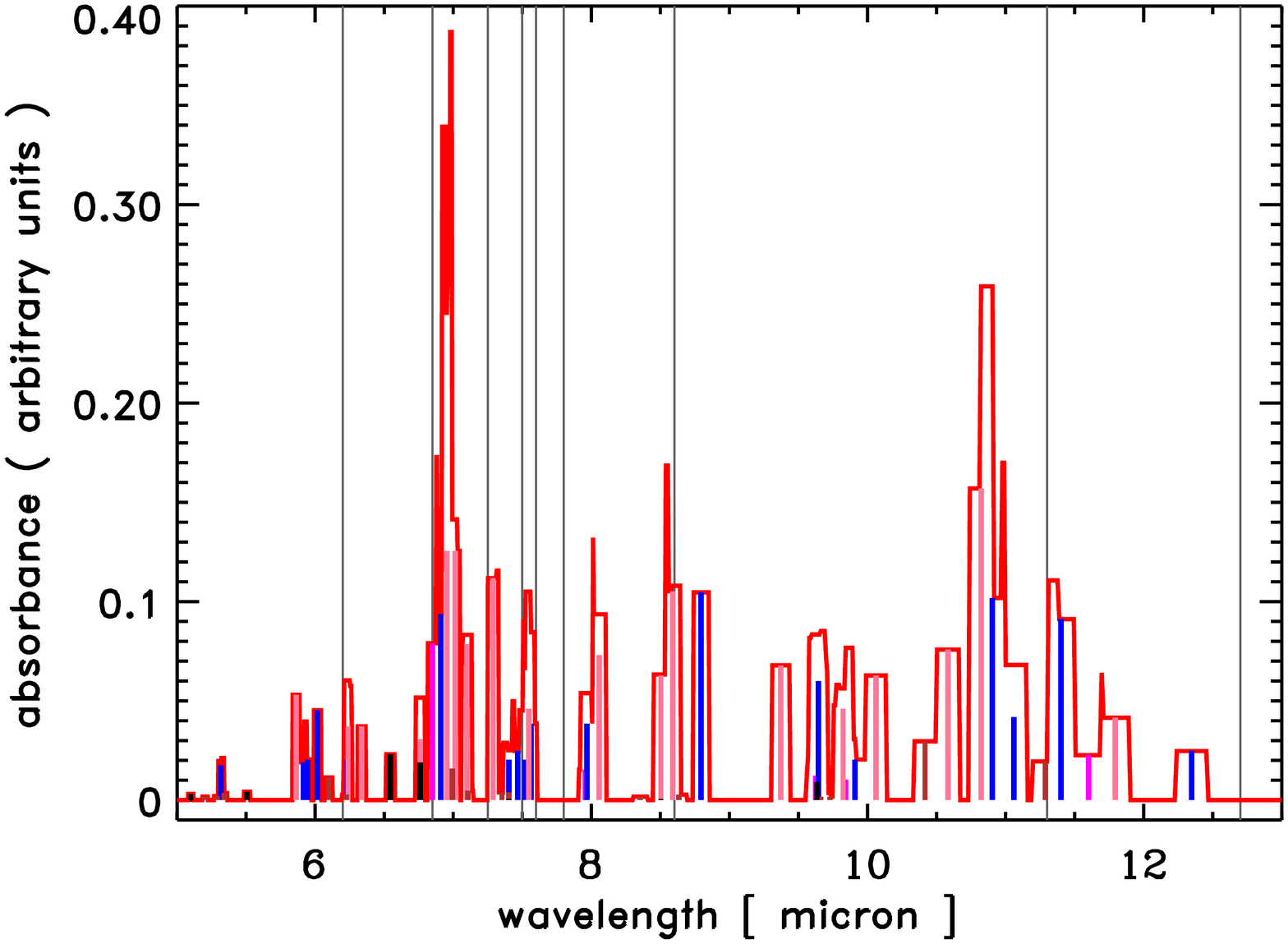} \\[-0.5cm]
\caption{The red line a multi-cyclic species absorbance spectrum as a linear combination of the six spectra (in the same order as above), in this case in the ratio 
0.14 : 0.27 : 0.41 : 0.03 : 0.03 : 0.14 
and schematically illustrates what a mixed component spectrum could look like in the $3\,\mu$m range (upper) and over the $5-13\,\mu$m wavelength region (lower). The vertical coloured lines show (same colour scheme as in the previous figure) show the positions and relative intensities of the six contributing molecules. 
The vertical dark grey lines indicate the central positions of the observed IR emission bands and the thin grey lines mark the positions of bands in laboratory-measured samples, as described \citep[][and references therein]{Bouteraon_etal_2018}. The light blue lines indicate the positions of the IR bands observed in absorption towards the Galactic Centre \citep{2002ApJS..138...75P}.}
\label{fig_stick_spectra}
\end{figure}

Fig.~\ref{fig_stick_spectra} shows the same data as in Fig.~\ref{fig_cyclics_spectra} but presented in absorbance in the form of a "stick" spectrum (coloured lines, same molecule colour code as in Fig.~\ref{fig_cyclics_spectra}). These data are overlain by a "smoothed" version of these summed data in order to schematically illustrate the combined spectrum in the $3.2-3.7\,\mu$m and $7-13\,\mu$m wavelength windows (red lines) in order to make a general comparison with observations. These figures taken together show that there are no strong bands that are inconsistent with the observed mid-IR emission bands (vertical dark grey lines). In particular, and worthy of note are the two concentrations of bands in the $6.5-7.5\,\mu$m and $9.5-10.5\,\mu$m wavelength regions, which would seem to be consistent with the very broad underlying bands, peaking at $\simeq 6.9\,\mu$m and $\simeq 10.0\,\mu$m, seen in decompositions of PDR dust emission spectra \citep[{\it e.g.},][]{2001A&A...372..981V}. 

In conclusion, it would appear that it is time to re-evaluate the origin of the IR emission bands and to accept that completely aromatic species (PAHs) are really not the best model to interpret the observed emission spectra, especially in the $\lambda = 3-4\,\mu$m region. 
It would seem that more complex, mixed aliphatic/olefinic/aromatic structures are likely to be a much better model to interpret these spectral features and should therefore be explored in more detail both experimentally and theoretically. 



\newpage
\noindent {\rhfont \Large 4.2. The spectral evolution of cyclic aliphatics and olefinics in the ISM}\\[-1.1cm]
\subsection{The spectral evolution of cyclic aliphatics and olefinics in the ISM} 
\label{sect_spect_evol}

Given the large spectral variations that aliphatic, olefinic and aromatic molecules exhibit in absorption in the $3-4\,\mu$m wavelength region it is perhaps surprising that the observed interstellar absorption and emission bands show so little variation. However, there is a clear and distinct difference between the $3-4\,\mu$m bands seen in emission and absorption in the ISM (see lower panel in Fig.~\ref{fig_cyclics_spectra}). The major difference being that the $3.30\,\mu$m band, which is dominant in emission, is hardly evident in the dust absorption seen towards the Galactic Centre. Also, the $3.40\,\mu$m band in emission appears at $3.42\,\mu$m in absorption. 

Here we qualitatively analyse the absorption spectra, as measured by transmission/absorbance, but note that the spectra of such cyclic hydrocarbon moieties within larger particles will, in emission, likely exhibit different band-to-band ratios, which depend upon the particle structure and the source of excitation. 
For these six cyclic molecules, Table~\ref{table_cyclo_struc} shows the effect on the number of CH$_n$ groupings that would result if the equivalent cyclic structures were part of an a-C(:H) network. In this case the three network linking bonds are assumed to be trigonally-coordinated as shown in Fig.~\ref{fig_cyclics}. 
The first thing that we can see from the 20 entries in Table~\ref{table_cyclo_struc} is that there is a net reduction, by about a factor of two, in the number of CH$_n$ groupings for 12 table entries, for 3 (1) there is no change (an increase), and in 4 cases 3$^\circ$ CH bonds are formed where none exist in the parent molecule. Thus, it is to be expected that the spectra of polycyclic aliphatic/olefinic moieties within an a-C(:H) network will be systematically different from the isolated molecules. For instance, and as a close look at Table~\ref{table_cyclo_struc} will show, the $3.29$ and $3.51\,\mu$m ($3.46\,\mu$m) bands would be expected to decrease (increase) in strength, while the $3.40$ and $3.56\,\mu$m bands would be expected to be little affected. 
The net effect of the expected spectral differences between the isolated cyclic molecules and their network moiety variants would to bring the somewhat poor correspondence between the observed and modelled spectra shown in Fig~\ref{fig_cyclics_spectra}, in the $3.4-3.6\,\mu$m region, into better qualitative agreement.  
Nevertheless, the $3.29\,\mu$m band, as shown here in absorption, is weak compared to the $3.4\,\mu$m bands. However, it should be noted that in emission the band ratios will depend on the exact structures and, perhaps more importantly, the source of excitation and hence on the details of the emission process \citep{2012A&A...542A..98J}. 


In the outer regions of molecular clouds and towards the Galactic Centre it appears that the carbonaceous dust, seen predominantly in absorption, is dominated by aliphatic-rich materials \citep[{\it e.g.},][]{2002ApJS..138...75P,2012A&A...537A..27G,Faraday_Disc_paper_2014,2016A&A...588A..43J,2016A&A...588A..44Y}, whereas in the low density ISM and in PDRs, where the dust is seen predominantly in emission, the dust appears to be more aromatic-rich \citep[{\it e.g.},][]{2012A&A...542A..98J,2012ApJ...760L..35L,2013ApJ...770...78C,2013A&A...558A..62J}. Thus, in the transition from low to high excitation regions the form of the $3-4\,\mu$m spectra of the dust should evolve schematically in a top to bottom sequence following the general spectral characteristics indicated in Fig~\ref{fig_cyclics_spectra}. In essence, in low excitation regions $3.4$ and $3.5\,\mu$m bands will dominate and will be joined by and become subordinate to the $3.3\,\mu$m band in high excitation regions. 
However, the problem with observing such a spectral transition is that this change likely occurs extremely fast in the transition from low to high excitation (high to low density) regions. As has been shown \citep{2012A&A...542A..98J,2013A&A...558A..62J,Faraday_Disc_paper_2014,2015A&A...581A..92J}, the photo-processing time-scale is of the order of only thousands to hundreds of years in PDRs and must therefore occur over very short spatial scales making it hard to discern the spectral signatures of a-C(:H) dust transformation from aliphatic-rich (a-C:H) to a mixed aliphatic/olefinic/aromatic material (a-C). 
Nevertheless, there is some hope that with the launch and operation of the James Web Space Telescope (JWST), with its high spectral and spatial resolution, that we might finally be able to resolve this key stage in the evolution of carbonaceous dust in the ISM. 

In the observation of interstellar infrared emission spectra we are always and unavoidably averaging over many different contributing species and so we can never see the contribution from any single type of grain composition or its characteristic sub-structures. Further, and because we integrate across different environments with different physical conditions and therefore with different dust compositions, we may be seeing a mix of emission and absorption along a given line of sight. In addition, the dust structure also varies with the particle size \citep{2012A&A...542A..98J} leading to even more complexities that need to be taken into account in data in the course of the interpretation and modelling of observational data.

\vspace*{1.5cm}
\noindent {\rhfont \Large 5. Interstellar mantle formation and coagulation}\\[-1.1cm]
\section{Interstellar mantle formation and coagulation}  
\label{sect_mantles}


As dust transits from the diffuse to the dense ISM it undergoes evolution, which includes an increase in both the dust mass and the mean particle size, as species accrete from the gas phase to form mantles and as grains coagulate among themselves \citep[{\it e.g.},][]{2003A&A...398..551S,2009A&A...502..845O,2011A&A...532A..43O,2012A&A...548A..61K,2015ARA&A..53..541B}. 
Current studies tend to concentrate on the effects of coagulation with contemporaneous accretion usually neglected for reasons of computational convenience. Further, most coagulation studies consider only mono-modal dust size distributions.
More recent studies clearly indicate the importance of considering coagulation and mantle accretion in conjunction \citep{2015A&A...579A..15K,2016A&A...588A..43J,2016A&A...588A..44Y} and have tried to estimate the relevant time-scales involved \citep{Faraday_Disc_paper_2014}.  However, a detailed numerical study of the time sequence of coupled accretion and coagulation, including a fully-realistic dust size distribution, has yet to made. Nevertheless, and as a look at Fig.~5 in our earlier work shows \citep{Faraday_Disc_paper_2014}, the time-scale for carbon atom accretion from the gas, likely the predominant dust growth agent \citep{2015A&A...579A..15K,2016A&A...588A..43J,2016A&A...588A..44Y}, is a factor of $\simeq 3$ faster than the time-scale for nano-particle coagulation onto larger grains. Thus, it appears that accretion can, at least in the very earliest phases of dust evolution in the transit to denser regions, occur in the absence of coagulation but that coagulation does not occur without accretion. In fact, the accretion of mantles could actually aide coagulation by providing sticky surface coatings (initially carbonaceous and then icy) on all grains but most notably, and most effectively, upon the more refractory amorphous silicate grains. 
An effect that has indeed already been noted for icy surfaces \citep{1993ApJ...407..806C}.  

\begin{figure}[!h]
\centering\includegraphics[width=4.0in]{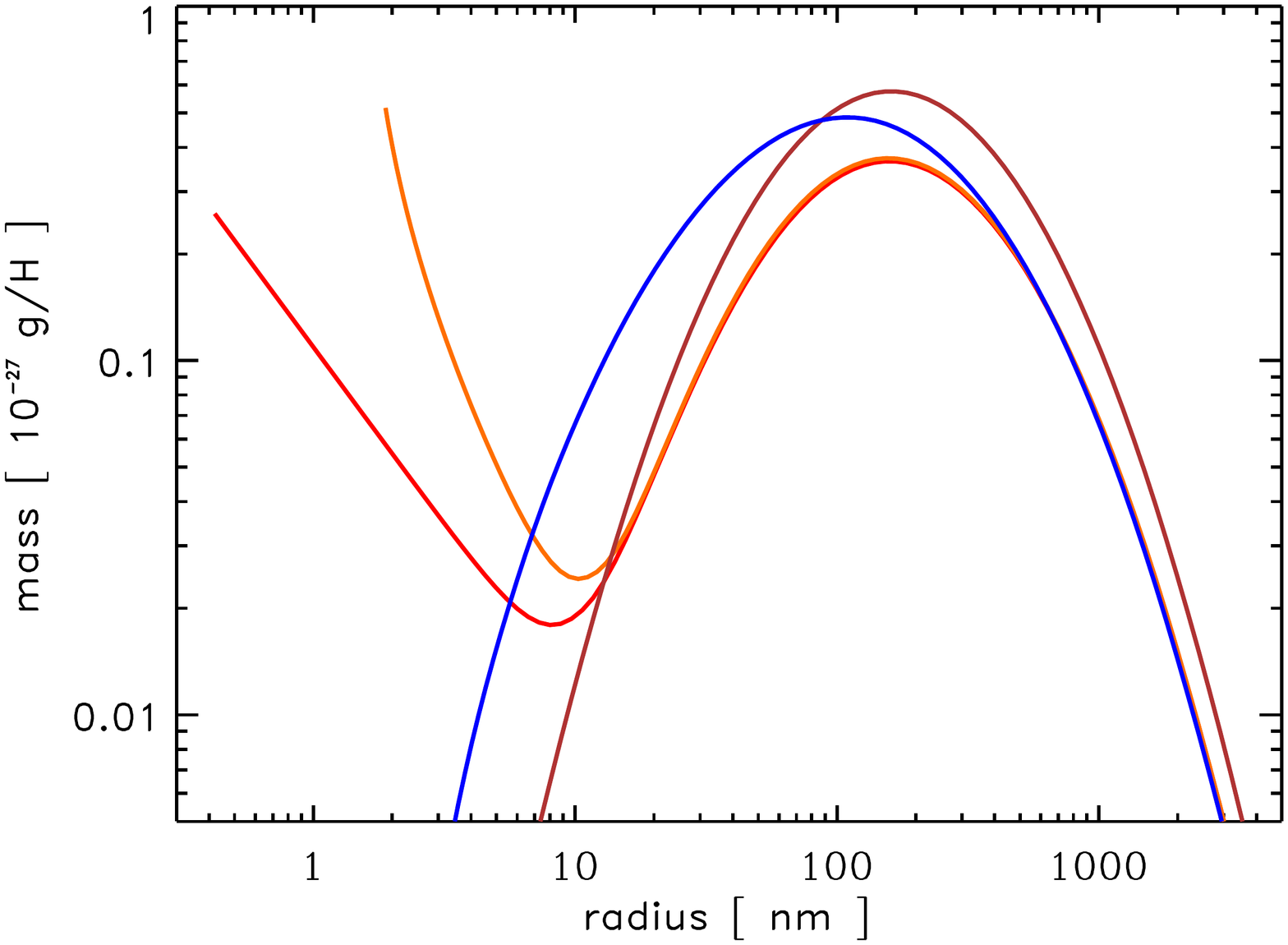}\\[-1.2cm]
\centering\includegraphics[width=4.0in]{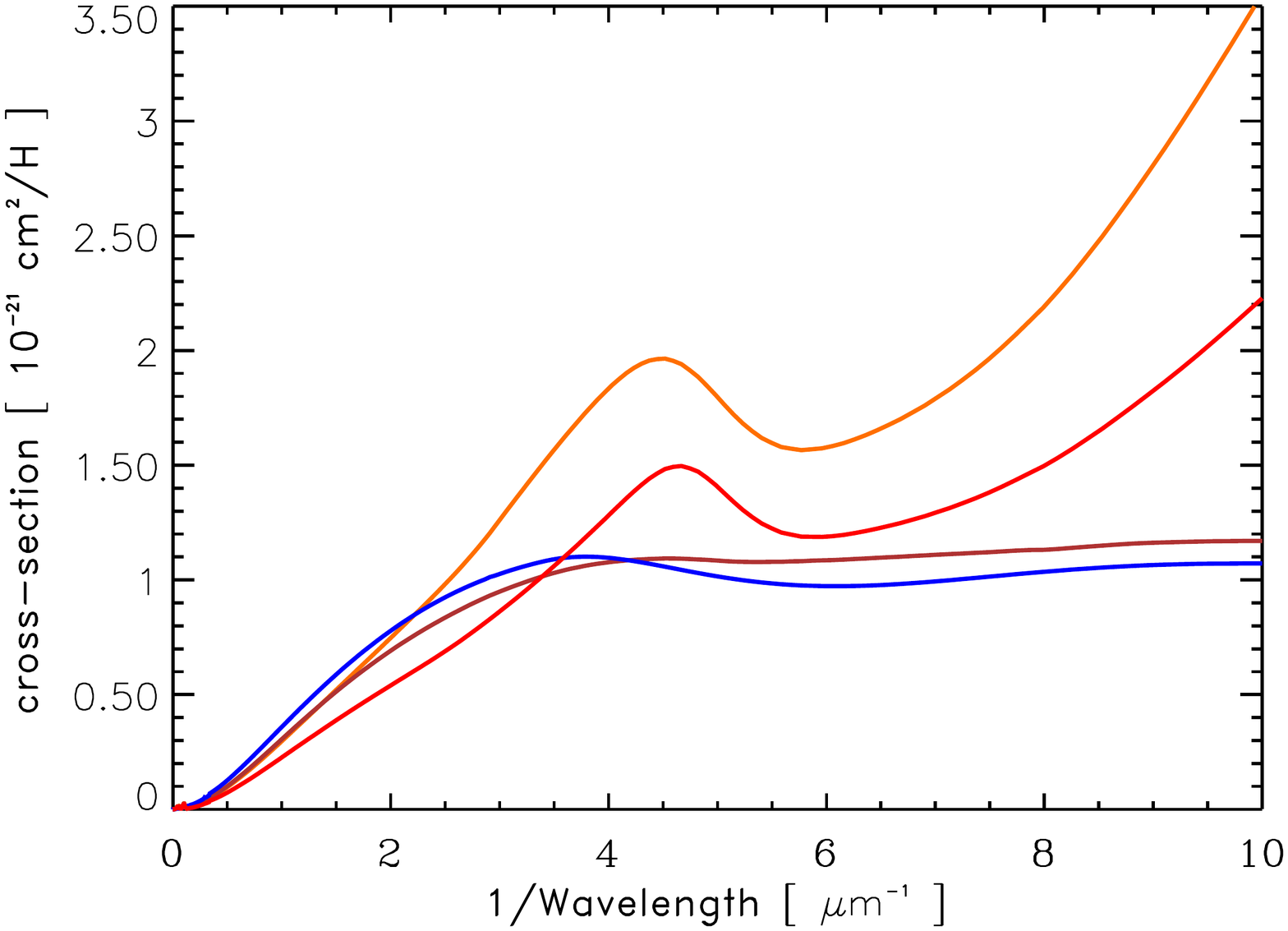} \\[-1.2cm]
\centering\includegraphics[width=4.0in]{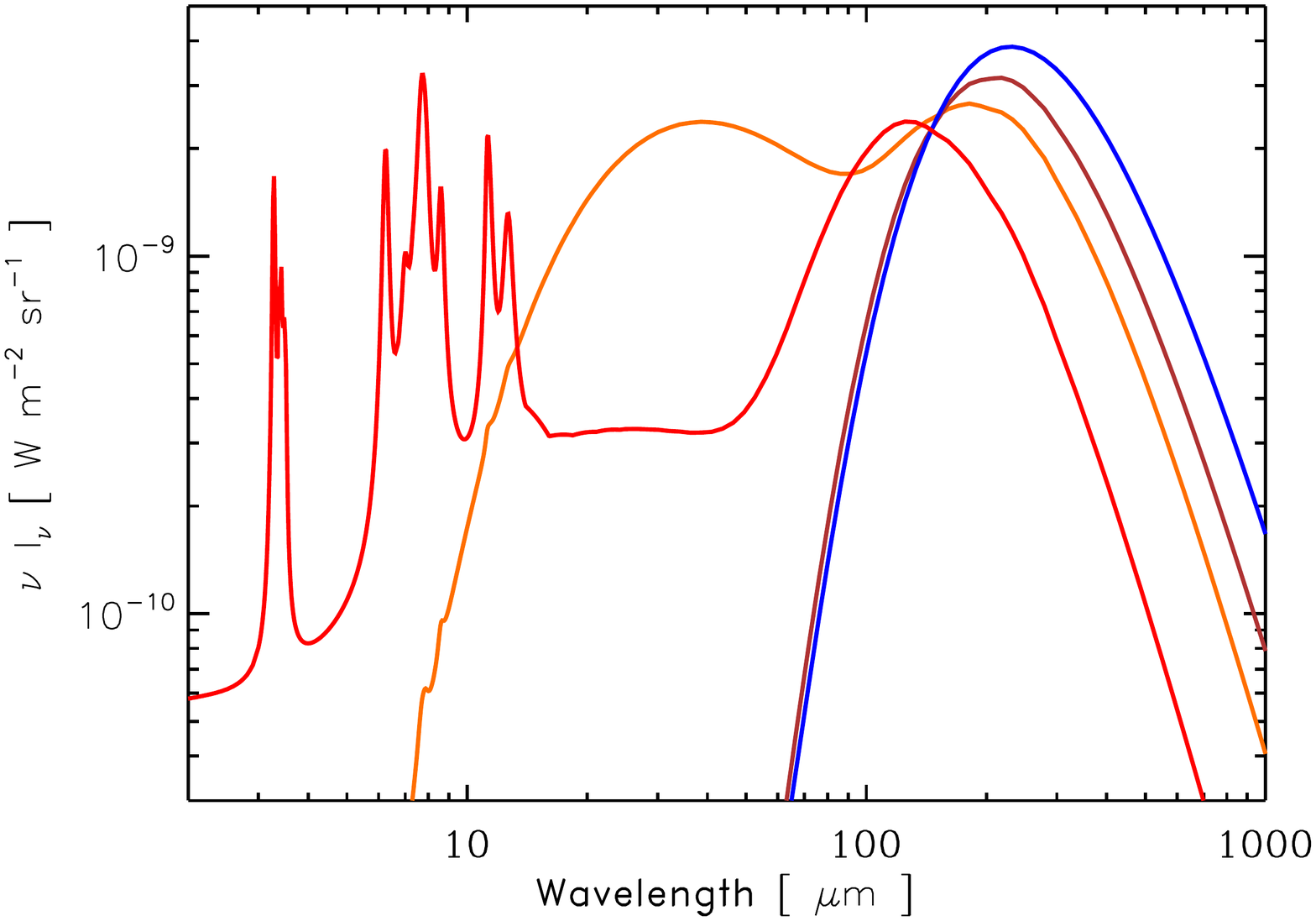} \\[-0.8cm]
\caption{A schematic view of dust evolution for the standard THEMIS diffuse ISM (red), THEMIS with the addition of 1.5\,nm thick mantles of a-C(:H) on all grains (orange), THEMIS a-C and a-C/a-C:H grains coagulated onto the a-Sil$_{\rm Fe,FeS}$/a-C grains (brown) and the combined effects of both accretion and coagulation (blue): size distributions (top), extinction (middle) and dust thermal emission (SED, lower). In the lower panel the standard diffuse ISM dust emission (red) is divided by ten to allow a better comparison with the shapes of the other SEDs, which assume an interstellar radiation field one tenth of that in the diffuse ISM.}
\label{fig_dust_evolution}
\end{figure}

%

It is possible to schematically show the effects of accretion and coagulation, as shown in Fig.~\ref{fig_dust_evolution} for the THEMIS dust model. In this case the accretion of the available gas phase carbon not already within the dust \cite[$\simeq 200$\,ppm, assuming a total cosmic abundance for carbon of 400\,pm,][]{2017A&A...602A..46J} will mantle all grain surfaces to the same depth ($\sim 1.5$\,nm). The a-C(:H) mantle depth is calculated using an instantaneous accretion approximation, {\it i.e.}, all the available gas phase carbon is accreted in one time-independent step. In order to calculate the effects of coagulation we also use an instantaneous approximation, assuming that all of the carbonaceous grains ({\it i.e.}, the THEMIS a-C nano-particles and core/mantle a-C/a-C:H grains) coagulate time-independently onto the surfaces of the carbon-mantled silicate grains ({\it i.e.}, the THEMIS a-Sil$_{\rm Fe,FeS}$/a-C grains). In this case we use a normalised coagulation matrix to apportion the carbon grain mass across the size distribution, either with or without accreted a-C(:H) mantles, {\it i.e.}, 
\[
\Theta^\prime_{ij} = n_i  \, n_j \, \pi ( a_i + a_j)^2 \, v_j \, S_{ij} 
\]
\begin{equation}
\Theta_{ij} =  \frac{ \Theta^\prime_{ij} }{ \sum_i  \sum_j  \Theta^\prime_{ij} }
\label{eq_coagulation}
\end{equation}
where $n_i$ and $n_j$ are the abundances of the $i^{\rm th}$ and $j^{\rm th}$ grains, $\pi ( a_i + a_j)^2$ is the collision cross-section, $v_j$ is the brownian motion velocity of the $j^{\rm th}$ grain and $S_{ij}$ is the coagulation efficiency (taken as unity and assumed to be independent of the grain size). With this approach $\Theta_{ij}$ gives the fraction of the total carbon grain mass (in both a-C and a-C/a-C:H grains) that is added to the $j^{\rm th}$ silicate grains ({\it i.e.}, a-Sil$_{\rm Fe,FeS}$/a-C) in the instantaneous coagulation approximation. This approximation method allows for a schematic comparison of the effects of accretion and coagulation on the dust size distribution as it evolves form low to high density regions. 

Clearly accretion, in the absence of any coagulation (orange lines in Fig.~\ref{fig_dust_evolution}), leads to an increase in the dust mass, which is most marked for the smallest grains, leading to a shift in their minimum size and a steepening of their size distribution. On the other hand, coagulation in the absence of accretion (brown line in Fig.~\ref{fig_dust_evolution}), leads to the loss of all small grains and a significant increase in the large grain dust mass. These effects are combined for coupled accretion and coagulation, leading to approximately log-normal size distributions (blue line in Fig.~\ref{fig_dust_evolution}). Given the current lack of appropriate optical properties for the optical properties of the multilayer, coagulated grains in even this simple model, it is not yet possible to undertake detailed calculations of the resultant dust extinction and emission. Nevertheless, and based on our earlier work \citep{2013A&A...558A..62J,2012A&A...548A..61K,2015A&A...579A..15K}, we schematically show the effects of the evolved dust size distributions on the dust extinction and emission in the middle and lower panels of Fig.~\ref{fig_dust_evolution}. In this case the standard diffuse ISM dust materials were used to illustrate the effects. However, and as we have already shown \citep{2016A&A...588A..43J,2016A&A...588A..44Y} dust evolution also has a significant effect upon the dust materials and their optical properties, which are not taken into account in this schematic view. 
Nevertheless, we can make some general statements about what might be expected, based upon the results shown in Fig.~\ref{fig_dust_evolution}. 
Firstly, the effects of accretion alone manifest as increased IR-UV extinction, steeper FUV extinction, a broader UV bump and increased dust emission in the mid-infrared (MIR) leading to broader and flatter SEDs \citep{2013A&A...558A..62J}.  
Secondly, the effects of coagulation, and coagulation combined with accretion, lead to a grey visible to UV extinction, decreased MIR emission and a slightly broader FIR emission peak in the SED \citep{2015A&A...579A..15K}. However, the width of the FIR dust emission peak will be dependent upon the optical properties used and so these schematic data should be regarded with caution here.  
Thus, if the onset of accretion occurs before coagulation, as shown in our previous study \citep{Faraday_Disc_paper_2014}, this will be reflected in the disappearance of the IR emission bands accompanied by a broadening of the SED and increased MIR continuum emission. Then once coagulation starts to kick-in the MIR continuum will drop away and the dust emission will look increasingly blackbody-like with a single SED peak in the FIR. Further, it would appear that the dust size distribution in the densest regions of the ISM and in circumstellar shells is most probably best represented by a log-normal distribution rather than the widely-used power laws. However, wherever nano-particles are present in abundance they probably do exhibit power-law size distributions due to their formation through some kind of photo-processing-induced or collisional fragmentation process.  
These changes in the dust size distribution, and hence in the dust extinction and the dust thermal emission (SED), appear to be in general agreement with observations \citep{2013A&A...558A..62J,2015A&A...579A..15K,2016A&A...588A..43J,2016A&A...588A..44Y}. 

\vspace*{1.5cm}
\noindent {\rhfont \Large 5.1. Evolving gas-to-dust ratios}\\[-1.1cm]
\subsection{Evolving gas-to-dust ratios}  
\label{sect_GtoD}

Given that the THEMIS modelling framework encompasses dust evolution as the dust transits between interstellar clouds environments this engenders an accompanying change in the gas-to-dust mass ratios (G/D) with environment, Table~\ref{table_GtoD} summarises the THEMIS G/D values. In the transition between diffuse and denser ISM regions \citep{2012A&A...548A..61K,2015A&A...579A..15K}, where basically all available matter becomes accreted in the form of carbon-rich, carbon/oxygen-rich and ice mantles, with increasing density \citep{2016A&A...588A..43J,2016A&A...588A..44Y,2016RSOS....360224J}. This evolutionary sequence is indicated by the upper four entries in Table~\ref{table_GtoD}. In this case a decrease in G/D from 135 to 55 is clear implying that, for a  galaxy with a Milky Way type metallicity, the minimum achievable G/D value is of the order of $50-60$.  We now consider dust evolution in the opposite sense, {\it i.e.}, from low excitation diffuse clouds to high energy regions such as active star formation regions, photo-dissociation regions (PDRs) and supernova-generated shocks, in order of approximately increasing energy density. In this sequence the G/D values increase dramatically, tending towards the extremely high G/Ds observed typical of most dwarf irregular galaxies \citep{2014A&A...563A..31R}.

\begin{table}[!b]
\caption{The gas-to-dust mass ratios (G/D), dust mass relative to hydrogen, dust mass relative to the available metals, carbon and oxygen abundances, [C] and [O], in dust (in parts per million, ppm) and the percentage by volume of carbonaceous matter in dust, $V_{f,{\rm C}}$. This is shown as a function of the ISM environment and the corresponding dust model, where: DISM indicates the standard diffuse ISM dust model, C (O) carbon (oxygen) atom accretion from the gas and ice the presence of ice mantles. Fractional variations from the standard diffuse ISM abundances of carbonaceous nano-particles (big grains) $\frac{1}{n}$C$_{\rm np}$ ($\frac{1}{n}$C$_{\rm bg}$) or no contribution at all (0$\times$). Low Z indicates sub-solar, low metallicity environments.} 
\label{table_GtoD}
\begin{tabular}{lcclcccccc}
\hline
                     &   $\approx n_{\rm H}$  &  $\approx T_{\rm gas}$   &            &        & $M_{\rm dust}$  &  dust    & [C]      & [O]     & $V_{f,{\rm C}}$ \\[-0.2cm]
                     &     &  &            &        &     ----------          &  --------- &            &           &                         \\[-0.2cm]
Environment &  (cm$^{-3}$) &  (K) & Dust type & G/D & $m_{\rm H}$      & metal   & (ppm) & (ppm) & (\%)                 \\
\hline
dense          &  $10^4$ & 15 &  DISM+C+O+ice                                                       &   55 & 0.0184 & 0.88 & 406 & 566 & 41 \\
translucent  &   1500    & 20 &  DISM+C+O                                                             &   81 & 0.0124 & 0.60 & 406 & 270 & 74 \\
translucent  &   1500    & 20 &  DISM+C                                                                  &  102 & 0.0098 & 0.47 & 406 & 110 & 65 \\
diffuse         &     50      & 100 &  standard DISM                                                        &  135 & 0.0074 & 0.36 & 206 & 110 & 47 \\
diffuse         &    50       & 100 &  $\frac{1}{2}$C$_{\rm np}$                                          & 153  & 0.0066 & 0.32 &135 & 110 & 36 \\
diffuse         &    50       & 100 &  $\frac{1}{10}$C$_{\rm np}$                                          & 170  & 0.0059 & 0.28 & 77 & 110 & 23 \\
diffuse         &    50      & 100 &  $\frac{1}{10}$C$_{\rm np}$, $\frac{1}{2}$C$_{\rm bg}$ & 180 & 0.0056 & 0.27 & 52 & 110 & 17 \\
energetic     &   0.25    & $10^4$ &  0$\times$C$_{\rm np}$, $\frac{1}{2}$C$_{\rm bg}$      & 185      & 0.0054 & 0.26 & 38 & 110 & 13 \\
energetic     &   0.25    & $10^4$ &  0$\times$C$_{\rm np}$, 0$\times$C$_{\rm bg}$        & 196  & 0.0051  &  0.25 & 13 & 110 & 5\\
energetic     &   0.25    &  $10^4$&  bare a-Sil                                                                    & 202  & 0.0049 & 0.24 & 0 & 110 & 0 \\
low Z/x-ray  &   0.01    &  $10^6$ &  $\frac{1}{3}$ a-Sil                                                       & 613   & 0.0016 & 0.08 & 0 & 36 & 0 \\
low Z/x-ray  &   0.01    &  $10^6$ &  $\frac{1}{30}$ a-Sil                                                      & 6742 & 0.0002 & 0.01 & 0 & 4 & 0 \\\hline
\end{tabular}
\end{table}


\begin{figure}[!h]
\centering\includegraphics[width=5.5in]{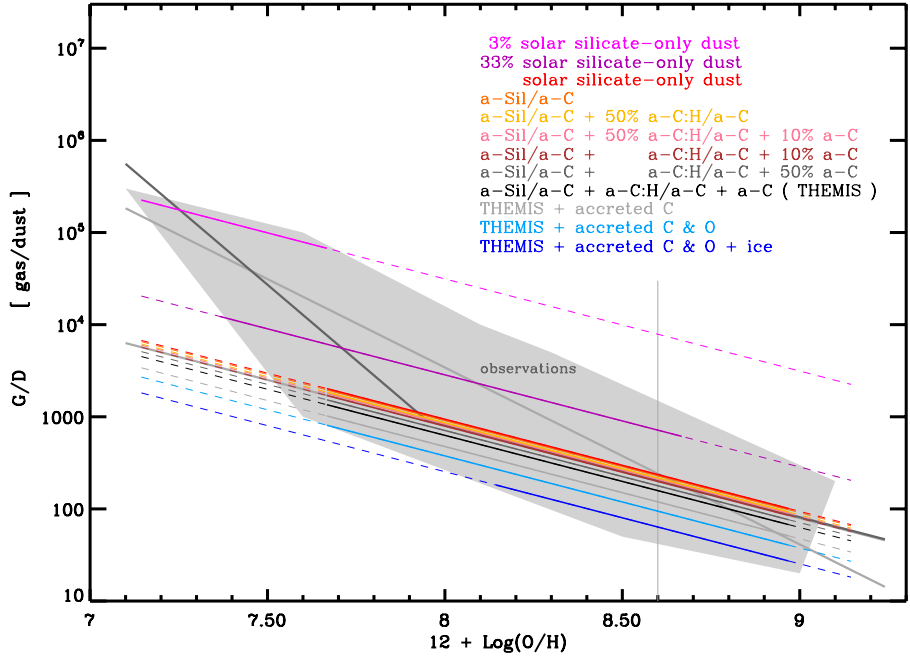}
\caption{THEMIS model gas-to-dust ratios (G/D) as a function of metallicity, expressed as 12+Log(O/H), and ISM phase from dense clouds (blue and light blue) to diffuse ISM (grey and black) to energetic regions where dust is either depleted or severely processed and eroded due to environmental effects (brown to violet). The grey shaded area shows the observed range of gas-to-dust ratios as a function of metallicity \citep[taken from][]{2014A&A...563A..31R}. The Milky Way metallicity is 12+Log(O/H) $\simeq 8.6$ and is shown by the vertical grey line.}
\label{fig_THEMIS_GtoD}
\end{figure}

\begin{figure}[!h]
\centering\includegraphics[width=5.5in]{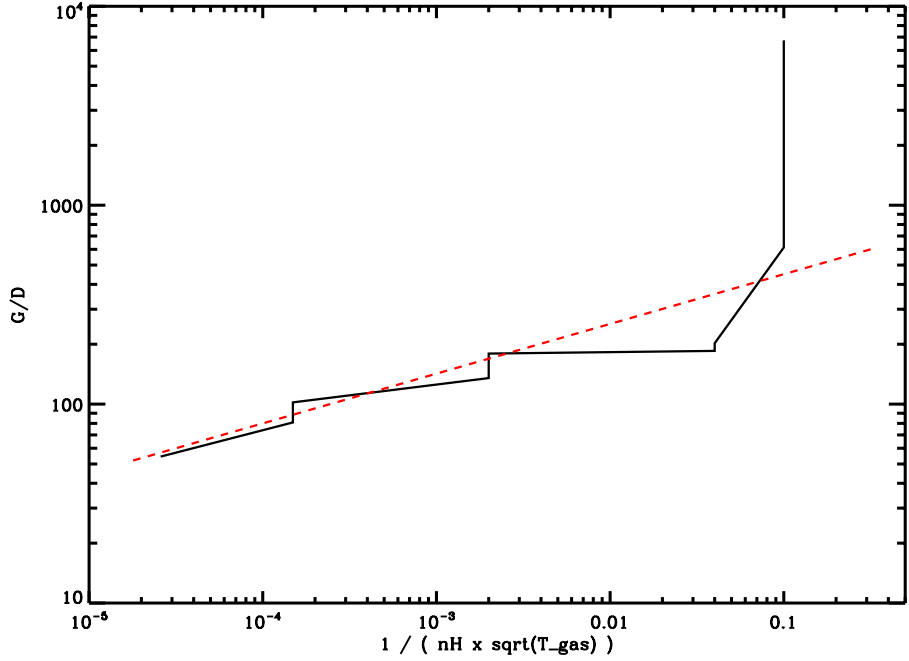}
\caption{THEMIS model gas-to-dust ratios (G/D) shown in Table~\ref{table_GtoD} plotted as a function of the accretion parameter $\xi = 1/( n_{\rm H} \surd T_{\rm gas})$. The dashed red line shows the empirical relationship ${\rm G/D} = 800 ( n_{\rm H} \surd T_{\rm g})^{-0.25}$.}
\label{fig_THEMIS_accn_param}
\end{figure}

How G/D varies with galaxy evolution and metallicity is a key constraint for understanding both galaxy and dust formation and evolution, both locally and through cosmic time
\citep{1990A&A...236..237I,1998ApJ...496..145L,2002A&A...388..439H,2002MNRAS.335..753J,2005A&A...434..849H,2007ApJ...663..866D,2008ApJ...678..804E,2008ApJ...672..214G,2009ApJ...701.1965M,2010MNRAS.402.1409B,2011A&A...532A..56G,2011A&A...535A..13M,2014A&A...563A..31R}. 
Fig.~\ref{fig_THEMIS_GtoD} schematically summarises the available observational data and the THEMIS predictions. In this figure the grey shaded area shows the G/D for a wide range of galaxy types covering a wide range of metallicities expressed in the usual way as $12+{\rm Log(O/H)}$ \citep{2014A&A...563A..31R}. In galaxy studies G/D variations are often interpreted in terms of a critical metallicity or dust formation threshold at around a metallicity of $12+{\rm Log(O/H)} = 8$ (see Fig.~\ref{fig_THEMIS_GtoD}), shown by the thicker dark grey line with a break in slope at close to this critical metallicity. The lighter grey line of constant slope shows the general linear trend that it is possible to draw through the cloud of galaxy data. Fig.~\ref{fig_THEMIS_GtoD} also shows the variation in the THEMIS model G/D ratios for the dust compositions shown in the figure inset and in Table~\ref{table_GtoD}. In this table and figure G/D is assumed to follow a constant dust to metals ratio for each composition, {\it i.e.}, the same fraction of metals are assumed to be tied up in dust for each given composition. As can be seen in Fig.~\ref{fig_THEMIS_GtoD}, the observations tell us that we do not see all possible dust compositions at all possible metallicities. The data-model comparison seems to indicate that there is a clear trend for dust typical of high energy regions at low metallicities. This would seem to indicate a strong role for environmental factors in explaining dust evolution through metallicity space. 

As many studies have shown \citep{2008ApJ...672..214G,2013EP&S...65..213A,2014A&A...562A..76Z}
dust formation in the ISM appears to be an unassailable requirement in explaining dust evolution in galaxies. Hence, a close look at the simple fundamentals of dust accretion in the ISM is merited. In essence the key accretion time-scale is given by
\begin{equation}
\tau_{\rm acc} = ( \,n_i \,\sigma_{\rm d} \,v_i \,s_i\,)^{-1} =  \Bigg\{\, n_{\rm H} \, X_i \,\pi a_{\rm d}^2 \left( \frac{2k_{\rm B} T_{\rm g}}{m_i} \right)^{\frac{1}{2}} s_i  \Bigg\}^{-1}
\end{equation}
or 
\begin{equation}
\tau_{\rm acc} = \frac{1}{\surd(2 \, k_{\rm B})} \left[ \, \frac{X_i}{m_i} \, s_i \,\pi a_{\rm d}^2 \right]^{-1} \ \Bigg\{ \, n_{\rm H} \, \surd T_{\rm g} \Bigg\}^{-1}
\label{eq_acc_param}
\end{equation}
where $n_i$, $v_i$, $s_i$, $X_i$, $m_i$  are the accreting gas phase species, $i$, abundance, velocity, sticking efficiency, abundance relative to H and mass. $\sigma_{\rm d}$ is a single dust grain geometric cross-section, $n_{\rm H}$ is the H abundance abundance in the gas, $a_{\rm d}$ the grain radius, $k_{\rm B}$ the Boltzmann constant and  $T_{\rm g}$ the gas temperature. 
The re-structuring of the accretion time-scale, in Eq.~(\ref{eq_acc_param}), separates the variables into dust-dependent and environment-dependent terms, the square and curly brackets, respectively. 
Under the supposition that dust evolution is weak or negligible, {\it i.e.}, that dust forms wherever and whenever possible and that the dominant dust size is fixed, the term in square brackets can be regarded as a constant, we can then compare the combination of the gas density and temperature, the environmental factors, for different interstellar regimes. Clearly, for highly excited regions, such as those with high G/D in Fig.~\ref{fig_THEMIS_GtoD}, this may no longer hold true but can be left aside for the purposes of the present exercise. 
We are now left with a factor, $1/( n_{\rm H} \surd T_{\rm g})$, which here will be called the environmental accretion parameter, or simply the accretion parameter $\xi$. In Fig.~\ref{fig_THEMIS_accn_param} G/D has been plotted as a function of the accretion parameter, $\xi = ( n_{\rm H} \surd T_{\rm g})^{-1}$, for the dust compositions shown in Fig.~\ref{fig_THEMIS_GtoD}, under the empirically-assumed gas densities and temperatures indicated in Table~\ref{table_GtoD}. This figure also shows a power-law approximation that generally follows the observed behaviour for G/D $< 10^3$. For the parameter space delimited in Table~\ref{table_GtoD} the accretion parameter $\xi$ spans about three orders of magnitude, from $2.6 \times 10^{-5}$ to $0.04$ in units of cm$^3$\,K$^{0.5}$, which is really a pseudo-pressure measure, but that $\xi^{0.25}$ ranges from only $0.07$ to $0.45$, {\it i.e.}, spans a factor of only 6 or 7. 

We therefore propose that G/D, which for galaxies is generally measured via the ratio of the observed atomic gas mass and a model-derived dust mass, $M_{\rm HI}/M_{\rm dust}$, can be expressed empirically, for a solar metallicity region or galaxy, as a function of the the proposed accretion parameter $\xi$ and that the behaviour can approximated as 
\begin{equation}
\frac{M_{\rm HI}}{M_{\rm dust}}  = \frac{G}{D}(n_{\rm H},T_{\rm g}) = \frac{800}{( \, n_{\rm H} \surd T_{\rm g} \, )^{\frac{1}{4}}} = 800 \, \xi^{\frac{1}{4}}. 
\end{equation}
The metallicity, $Z$, of a galaxy or region, can be expressed relative to solar metallicity, $Z_\odot$, {\it i.e.}, as ($Z/Z_\odot$), and so we can generalise the above expression for G/D to the form  
\begin{equation}
\frac{G}{D}(n_{\rm H},T_{\rm g},Z) = 800 \, \left( \frac{Z}{Z_\odot} \right) \, \xi^{\frac{1}{4}}. 
\end{equation}
It should be noted that this analytical approximation seems to imply only a weak dependence of dust formation in the ISM, by accretion from the gas, perhaps a posteriori confirming the initial assumption above that it is environment that primarily determines dust evolution rather than the nature of the dust itself. Or, to put it another way, dust forms whenever and wherever it can under the prevailing and local gas conditions. 
In this case, and as previously determined \citep{2011A&A...530A..44J}, the most physically-reasonable scenario is that the silicate/oxide dust component is the more refractory and resistant dust phase, except under the extreme conditions to be found in supernova remnants, and it is the carbonaceous dust phase that rather efficiently (re-)forms and (re-)accretes in the ISM under favourable outer molecular cloud and dense cloud conditions \citep[{\it e.g.},][]{2011A&A...530A..44J,Faraday_Disc_paper_2014,2015A&A...579A..15K,2016A&A...588A..43J,2016A&A...588A..44Y}. 

Clearly the simple empirical expression presented above fails for the highest values of G/D, {\it i.e.}, for G/D $> 10^3$, and only for the most extreme cases, which are likely to be regions of SN-shocked and/or x-ray emitting gas where dust is destroyed by high energy collisions with gas phase electrons and ions \citep[{\it e.g.},][]{1994ApJ...433..797J,1996ApJ...469..740J,2010A&A...510A..36M,2010A&A...510A..37M,2012A&A...545A.124B,2013A&A...556A...6B,2014A&A...570A..32B}. 

Consistent with the above dust evolution scenario, it needs to be emphasised that dust growth in the ISM probably always (and only) occurs in the densest regions. Thus, when we observe the dust in the very densest regions, such as cold cores, proto-planetary discs or PDRs, we are seeing it in a state that reflects the evolutionary processing that it underwent long before. In this sense, it makes no sense whatsoever to begin to model dust in discs with a power-law type of size distribution, which is often found to be typical of the low density diffuse ISM \citep[{\it e.g.},][]{1977ApJ...217..425M,1984ApJ...285...89D,2004ApJS..152..211Z,2007ApJ...657..810D,2014A&A...561A..82S}. 
Rather one should begin with something more akin to a log-normal distribution that is representative of dust that has been subject to accretion and coagulation in a preceding phase \citep{2012A&A...548A..61K,2015A&A...579A..15K}. 

Dust in the most numerous galaxies in the universe, the low metallicity dwarf galaxies, is an interesting case in point because they show systematically higher G/D than would be expected for their measured metallicities and this discrepancy in G/D can be more than an order of magnitude \citep{2014A&A...563A..31R} (see Fig.~\ref{fig_THEMIS_GtoD}).  The question is then, do these galaxies not form dust efficiently enough due to different circumstellar dust formation and evolution in low metallicity environments or is dust formation as efficient as elsewhere in other galaxies but that that dust, once formed, is then subjected to a hostile interstellar environment where it is rapidly destroyed?  This issue is still very much an open question and one that will require dedicated high resolution studies to resolve. Hopefully, the James Webb Space Telescope (JWST) will help to shed some light on this matter. 


\vspace*{1.5cm}
\noindent {\rhfont \Large 6. The interstellar and comet dust connection}\\[-1.1cm]
\section{The interstellar and comet dust connection}   
\label{sect_ISM_comets}

Perhaps rather surprisingly there appears to be a fundamental connection between the derived THEMIS gas to dust ratios (G/D)  and the dust composition and content of comets. A recent characterisation of the permittivity of comet 67P/CG with the CONSERT instrument onboard  Rosetta allowed a test of cosmochemical cometary dust models, principally consisting of silicate with various contributions of both insoluble and soluble carbonaceous materials \citep{2016MNRAS.462S.516H}. 
The results of this study indicate that cometary dust must contain a significant fraction of carbonaceous material in order to match the   CONSERT permittivity measurements. This work concluded that the required minimum carbonaceous material content is 75\% by volume;  about 2.5 times that of the CI chondrite Orgueil and significantly richer in carbonaceous matter than interplanetary dust particles, indicating that cometary dust contains a large reservoir of carbon \citep{2016MNRAS.462S.516H}. 

We now consider the upper four entries in the last column of Table~\ref{table_GtoD}, which shows $V_{f,{\rm C}}$, the volume fraction of carbon-rich material in each model mix. For the standard diffuse ISM G/D = 135 and 47\% of the total dust material volume is of carbonaceous matter. As we move to denser regions, {\it i.e.}, as we move up the last column from this entry, accretion onto grains will be dominated by carbonaceous matter \citep[see Section~\ref{sect_mantles} and ][]{2016A&A...588A..43J,2016RSOS....360224J,Faraday_Disc_paper_2014},  which serves to increase this fraction. 
Indeed as we pass from diffuse, through translucent and into denser regions (Table~\ref{table_GtoD}) the accreted mantles evolve in composition from a-C to a-C:H to a-C:H:O (O-rich a-C:H or a "carboxy" material) and finally to water ice-dominated mantles. 
The extra accreted carbon (a-C:H), and extra accreted carbon along with oxygen (a-C:H:O), is assumed to be in the form of low density a-C:H or ice-like materials (1.3 and $1.0$\,g\,cm$^{-3}$, respectively) and found to make up some 65 or $74$\% of the total volume. 
If, however, the a-C:H:O mantle material density is as high as $1.6$\,g\,cm$^{-3}$ then the carbonaceous matter volume fractions in dust would be reduced to 62\% and 71\% for a-C:H and a-C:H:O mantles, respectively, which therefore does not change the picture significantly. 
Thus, we find that the volume fraction of carbon-rich material in dense regions is $62-65$\% for extra accreted C and $71-74$\% for extra accreted C and O ("carboxy") mantle materials. This compares extremely well with observed value of 75\% for comet 67P, especially in the case of mixed "carboxy" mantles. 
In conclusion, it appears that the THEMIS dust evolution and modelling framework results are coherent with a widely-held view that cometary dust is a direct sampling of dense interstellar cloud dust materials. 

Comets are clearly the repositories of dense cloud interstellar matter that was mixed with solar nebula materials in the primitive proto-solar disc. The Rosetta/Philae mission to comet 67P/Churyumov-Gerasimenko has therefore given us a deeper insight into this primitive matter. 
It is therefore to be expected that the comparisons between interstellar and cometary dust likely goes much deeper than the overall carbonaceous matter volume content.  
The Cometary Sampling and Composition (COSAC) instrument detected a number of organic molecules with carbonyl ($>$C$=$O) bonds; including: 
aldehyde (R$-$C$\leqslant^{\rm H}_{\rm O}$),         
amide (R$-$C$\leqslant^{\rm NH_2}_{\rm O}$) and 
ketone $^{\rm R_{1}}_{\rm R_{2}}>$C$=$O              
functional groups, however, carboxylic acids, R$-$C$\leqslant^{\rm OH}_{\rm O}$, were not present in measurable quantities \citep{2015Sci...349b0689G}. About half of the COSAC-detected species contain carbonyl functional groups ($>$C$=$O and $=$C$=$O) and half contain nitrogen atoms in amine ($-$NH$_2$), imine ($-$N$=$) and nitrile ($-$C$\equiv$N) functional groups. 
Somewhat of a surprise though was that the most abundant molecule after water ($\sim 2$\% relative to water)  was formamide, H$-$C$\leqslant _{\rm O}^{\rm NH_2}$, with an O$=$C$-$N$<$ molecular configuration. 

It is also apparent that OCN configurations are rather common among the detected interstellar molecules and radicals containing at least three heavy atoms. For instance, and to date, among the detected species, this three-atom structure is found in at least five interstellar species, {e.g.}, cyanic acid (H$-$O$-$C$\equiv$N), isocyanic acid (H$-$N$=$C$=$O), formamide (H$-$C$\leqslant _{\rm O}^{\rm NH_2}$), methyl isocyanate (CH$_3$$-$N$=$C$=$O) and urea (O$=$C<$_{\rm NH_2}^{\rm NH_2}$), in various OCN configurations, {\it i.e.}, $-$O$-$C$\equiv$N, O$=$C$=$N$-$ and O$=$C$-$N. Further, related structures appear in fulminic acid (H$-$C$=$N$-$O) as O$-$N$=$C and in cyanoformaldehyde ($^{\rm H}_{\rm O}\hspace*{-0.05cm} \geqslant$C$-$C$\equiv$N) as O$=$C$-$C$\equiv$N configurations.   
A direct link between interstellar and cometary chemistry is therefore strongly implied. 

As per interstellar dust, cometary dust then represents matter that was, in major part, photolysed by stellar radiation into a resistant, amorphous, somewhat aromatic a-C carbonaceous solid. At the sizes of interstellar grains this material can hardly be regarded as refractory but at the size-scales of cometary surfaces this material would present a rather hard and resistant coating, far from anything ice-like. 
Thus, the surface of 67P is most probably reminiscent of a hard, photolysed, coal-like, carbonaceous matter and therefore it is in no surprise that Rosetta's Phil\ae\ lander was in for a hard landing and a bounce.
With a bit of deeper forethought, it would perhaps have been possible to foresee that latching onto the surface of a comet would be more like trying to "harpoon" a lump of coal than stapling a dirty snowball.  

\vspace*{1.5cm}
\noindent {\rhfont \Large 7. The importance of OCN molecules}\\[-1.1cm]
\section{The importance of OCN molecules}   
\label{sect_OCN}

This section follows up and and extends the discussion of OCN molecules introduced in the preceding section. 
After hydrogen and helium, oxygen, carbon, nitrogen and neon are by far the most abundant elements in the cosmos. Given that the chemistry of the noble gas elements He and Ne is rather limited, they will not be considered any further here. However, being the principal elements of organic chemistry and biological systems, H, O, C and N rightfully deserve some detailed attention. 

As a look at any list of the currently-detected interstellar and circumstellar molecules, radicals and ions will show, species with three or more different heavy atoms ({\it i.e.}, heavier than He) can be separated into two classes:
\begin{enumerate}
\item cyanides (--C$\equiv$N) and isocyanides (--N$\equiv$C) of metallic (Na, K, Mg, Al, Fe) or semi-metallic elements (Si), which are almost exclusively observed in the dense circumstellar shells around evolved stars such as IRC+10216) \citep[{\it e.g.},][]{2004A&A...426L..49G,2010ApJ...725L.181P,2011ApJ...733L..36Z,2013ApJ...775..133C,2017A&A...606L...5C}. 
\item species consisting of only H, O, C and N atoms\footnote{The molecule O=C=S, which is an analogue of CO$_2$, O=C=O, is here excluded as seemingly the only exception to this class.} which are observed in the dense regions of the ISM associated with star formation and with hot cores in particular. 
\end{enumerate}
In this section we are most interested in the second class and will no longer consider the more esoteric (iso)cyanide species. Among this second class of OCN species there is something rather special of note; namely, and in most cases, the three different heavy atoms are directly bonded to one another in a particular order, {\it i.e.}, in O$-$C$-$N conformations with varying bond order between them but almost always in this sequence and most often with a carbonyl (O$=$C$<$) termination (see above Section~\ref{sect_ISM_comets}), giving a seemingly rather fundamental O$=$C$-$N backbone structure. It seems curious that tri-heavy-atomic OCN molecules are not more diverse and that there should be so common a thread among so many observed species. Clearly, the chemistry of these elements will direct them to form such stable bond configurations. However, in the interstellar medium, where the elemental gas phase abundances are [O] $\approx 2 \times$[C] $\approx 2 \times$[N], why would the N atoms not be more widely-separated from the O atoms by more abundant C atoms in the detected species. One would, perhaps na\"{i}vely, have expected a much greater chemical variety in the molecules that have been observed. The observed OCN configurations (one molecule with $-$O$-$C$\equiv$N, two with O$=$C$=$N$-$ and three with O$=$C$-$N$<$ structures) therefore might appear to be rather fundamental and could perhaps be the result of a common formation pathway.  

Given that the O, C and N atoms (X) in the observed OCN species are conjoined, the carrier species must be rather hydrogen-poor. For example, as separated atoms in H$_2$O, CH$_4$ and NH$_3$ molecules the [H/X] ratio is 2.33, while [H/X] $= 1.67$ for HO$-$CH$_2-$NH$_2$, the most hydrogenated OCN structure possible. 
However, the least hydrogenated OCN structures O$=$C$=$N$-$H and H$-$O$-$C$\equiv$N, which are closer in nature to the observed interstellar species, have [H/X] ratios of only 0.33. For the detected interstellar OCN molecules, listed in the previous section, [H/X] ranges from 0.33 to 1.33, confirming that these molecules are rather hydrogen poor, which is paradoxical because they form and exist in the ISM a medium that is rich in hydrogen. 
The effects of hydrogenation will, in the low density ISM with its relatively hard interstellar radiation field ($\langle h\nu \rangle \sim 6-10$\,eV, {\it c.f.}, typical hydrocarbon bond energies of $3-9$\,eV), be counteracted by the effects of UV  photolysis which will tend to remove the more weakly bound hydrogen atoms from a molecule. Thus, the fact that interstellar molecules tend to be relatively hydrogen poor, while existing in a hydrogen-rich medium, strongly argues in favour of UV photolysis as the key player in their dehydrogenation, as appears to be the key element in the evolution of interstellar carbonaceous dust \citep{2012A&A...542A..98J,2013A&A...558A..62J,2014A&A...569A.119A,2015A&A...581A..92J,2015A&A...584A.123A}. 
A seemingly key role for UV photolysis also can be taken to imply that these molecules were actually formed in rather low density cloud environments, perhaps the translucent outer edges of molecular clouds, where a trade-off equilibrium formation pathway is achieved through a balance between the rather dense gas and moderate UV-shielding. Further, a critical role for chemically-actived (nano-)grain surfaces would appear to aid matters, with the accretion of a-C:H mantles on all grains providing a catalytic substrate for preferential and enhanced OCN species formation. We explore this possibility in the following section. 


\vspace*{1.5cm}
\noindent {\rhfont \Large 8. The relevance of epoxides to interstellar chemistry}\\[-1.1cm]
\section{The relevance of epoxides to interstellar chemistry}   
\label{sect_epoxides}

\begin{figure}[!h]
\centering\includegraphics[width=5.0in]{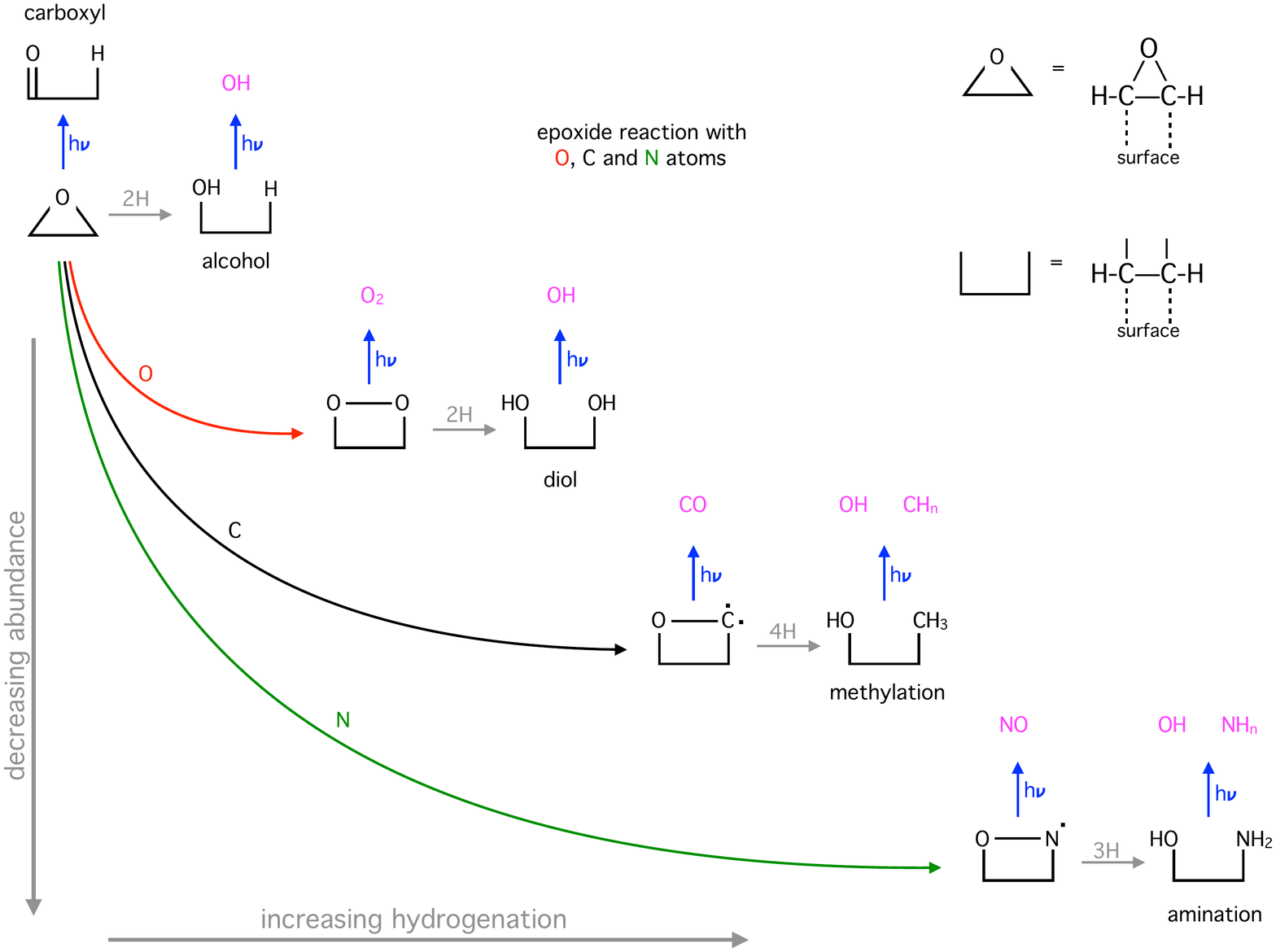}
\caption{Schematic examples of epoxide surface reactions with gas phase O, C and N atoms. Species shown in pink are likely gas phase products following UV photolysis.}
\label{fig_epoxide_1}
\end{figure}

\begin{figure}[!h]
\centering\includegraphics[width=5.0in]{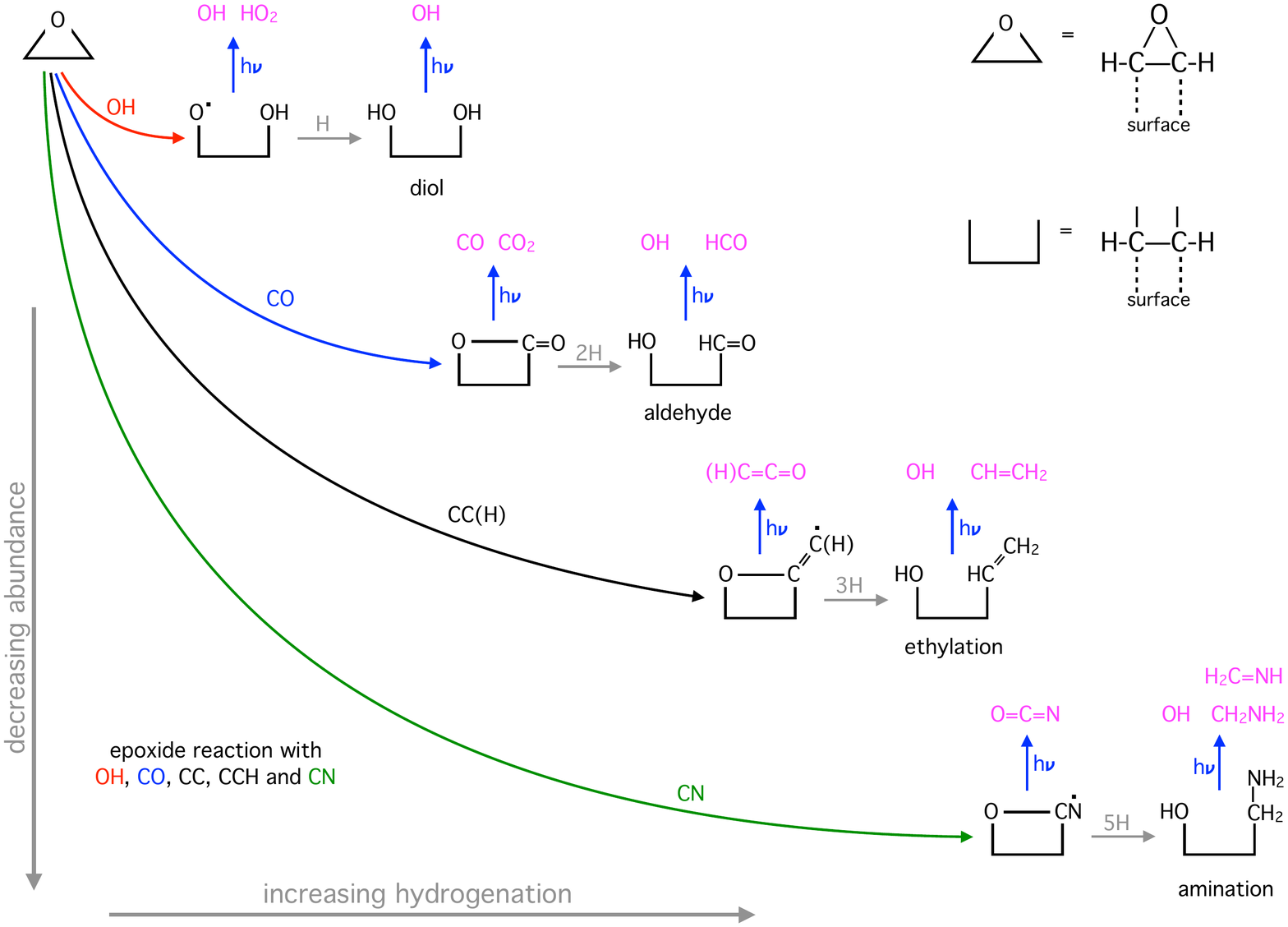}
\caption{Schematic examples of epoxide surface reactions with gas phase CO molecules and OH, CC(H) and CN radicals.}
\label{fig_epoxide_2}
\end{figure}

In previous work \citep{2016RSOS....360221J,2016RSOS....360224J} the possible role of epoxide-type ring (C$_2$O, $\triangle$) structures, >C$-\hspace{-0.24 cm} ^{\rm O}$C<, at the surface of interstellar carbonaceous grains was explored. The epoxide ring in ethylene oxide, also known as oxirane, is strained and chemically photo-reactive, with ring-opening reactions triggered by UV photons. Thus, epoxide structures resulting from the addition of O atoms to olefinic and aromatic C=C bonds in and on interstellar a-C(:H) grains could be rather abundant in the low-density ISM and could be the driver of some interesting and novel chemical reaction pathways to the observed interstellar molecules. Pathways that have not yet been explored within the field of interstellar chemistry \citep{2016RSOS....360221J,2016RSOS....360224J}. 
For example, epoxide structures (C$-\hspace{-0.24 cm} ^{\rm O}$C) would be expected to react with abundant gas phase atoms (H, O, C, N, \ldots), molecules (CO, \dots) and radicals (C$_2$, C$_2$H, CN, \ldots) etc. in a sequence of reactions that would depend upon their relative abundances. In the diffuse, low-density ISM the most likely epoxide reaction partners, in order of decreasing abundance, are H $\gg$ O, C, N > OH, CO > C$_2$, C$_2$H > CN and it is to be expected that all of these species will react with grain surface >C$-\hspace{-0.24 cm} ^{\rm O}$C< structures with little or no barrier to the reaction. Given that, epoxide ring-opening and reaction is triggered by UV photons, of which there is no lack in the ISM,\footnote{In the diffuse ISM the UV photon flux is of the order of $3 \times 10^7$\,photons\,cm$^{-2}$\,s$^{-1}$ \citep{2002ApJ...570..697H}.  Thus, a comparatively large 100\,nm radius interstellar a-C(:H) grain absorbs $\sim 10^{-2}$\,photons\,s$^{-1}$ or about one UV photon every few minutes and a nano-particle absorbs a UV photon approximately once every few weeks!} such reactions would seemingly be hard to avoid. 
Figs.~\ref{fig_epoxide_1} and \ref{fig_epoxide_2} illustrate the possible epoxide reaction pathways with the gas phase atoms O, C and N atoms and with gas phase CO molecules and OH, CC(H) and CN radicals leading to the 
very rich chemistry which is 
shown in Table~\ref{table_products}. 

From the lowest pathway in Fig.~\ref{fig_epoxide_2} and the last entry in Table~\ref{table_products} it would appear that a ready-made route, albeit subservient to many other possible branchings, to OCN species formation is likely through the reaction of relatively abundant gas phase CN radials with surface epoxide groups within the accreted and accreting a-C:H:O mantles. It is also possible that the reaction of N atoms with epoxide, following reaction with C atoms, from the gas or eroded from the grain surface, could also form and liberate OCN species from interstellar grain surfaces (lowest pathway in Fig.~\ref{fig_epoxide_1}). 

\begin{figure}[!h]
\centering\includegraphics[width=5.5in]{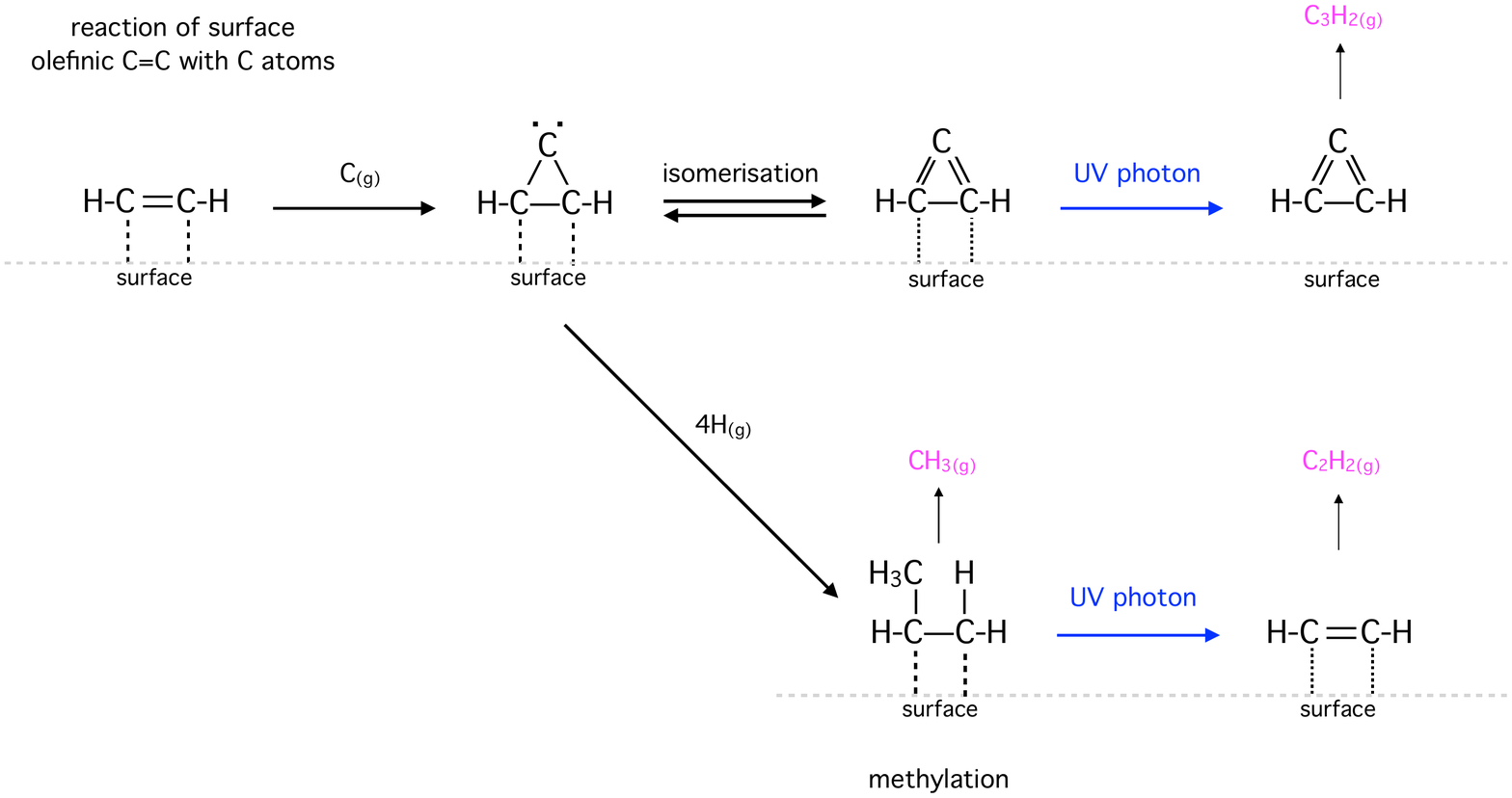}
\caption{Schematic examples of surface olefinic C=C reactions with gas phase C and H atoms and photolysis leading to grain surface erosion. Species shown in pink are likely gas phase products following UV photolysis.}
\label{fig_carbon_1}
\end{figure}

\begin{figure}[!h]
\centering\includegraphics[width=5.0in]{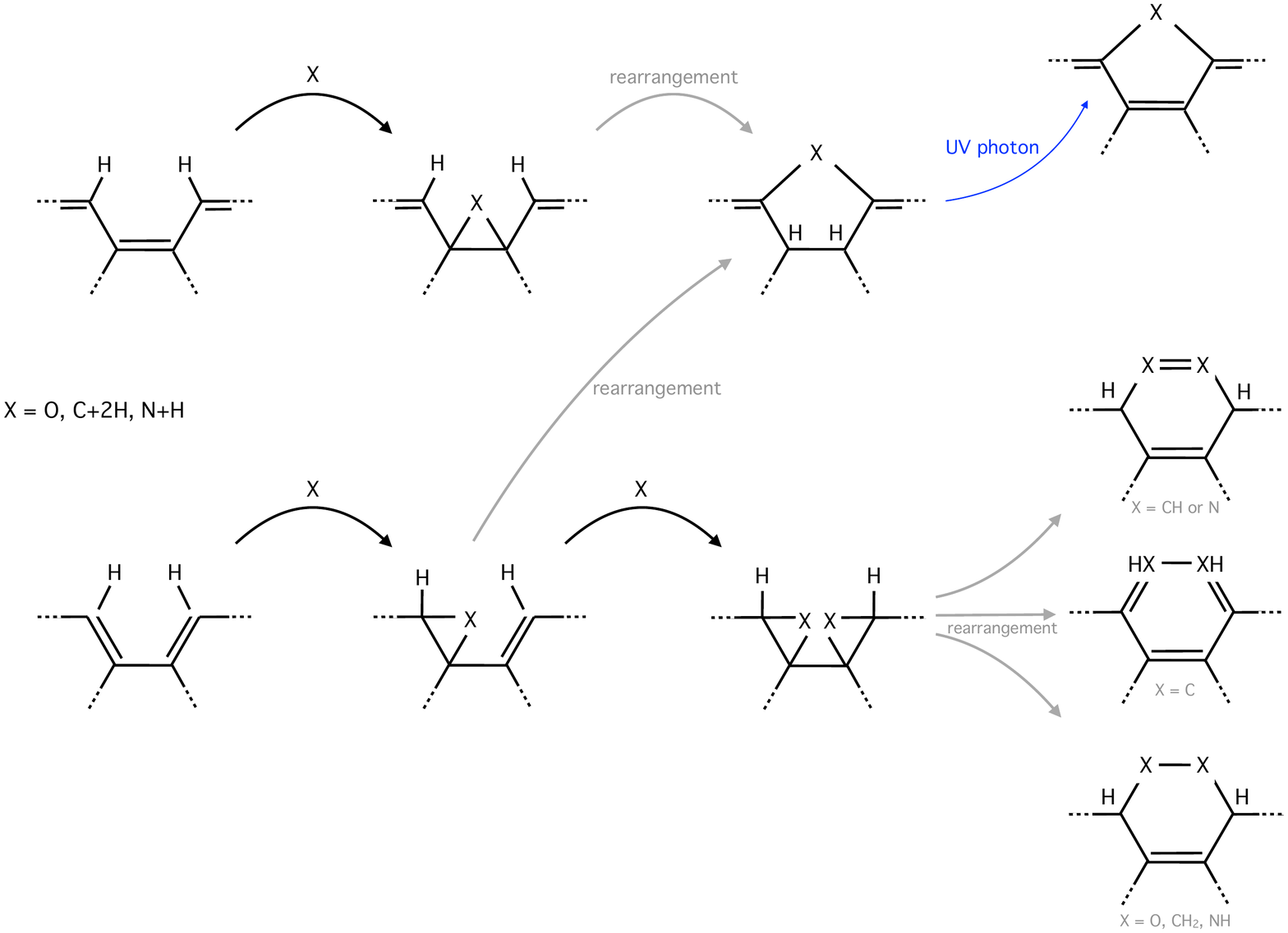}
\caption{Schematic surface grain growth ring-formation reactions with gas phase O, C(H$_{(2)}$), and N(H) radicals to form cyclohexene-, cyclohexadiene-, chromene- and (iso)quinoline-, pentacene-, furan-, (iso)indole- and carbazole-type homo- and hetero-cyclic ring structures, many of which are known colour centres and could therefore be DIB carrying sub-structures within a-C:H:X grains.}
\label{fig_ring_formation}
\end{figure}

\begin{figure}[!h]
\centering\includegraphics[width=5.0in]{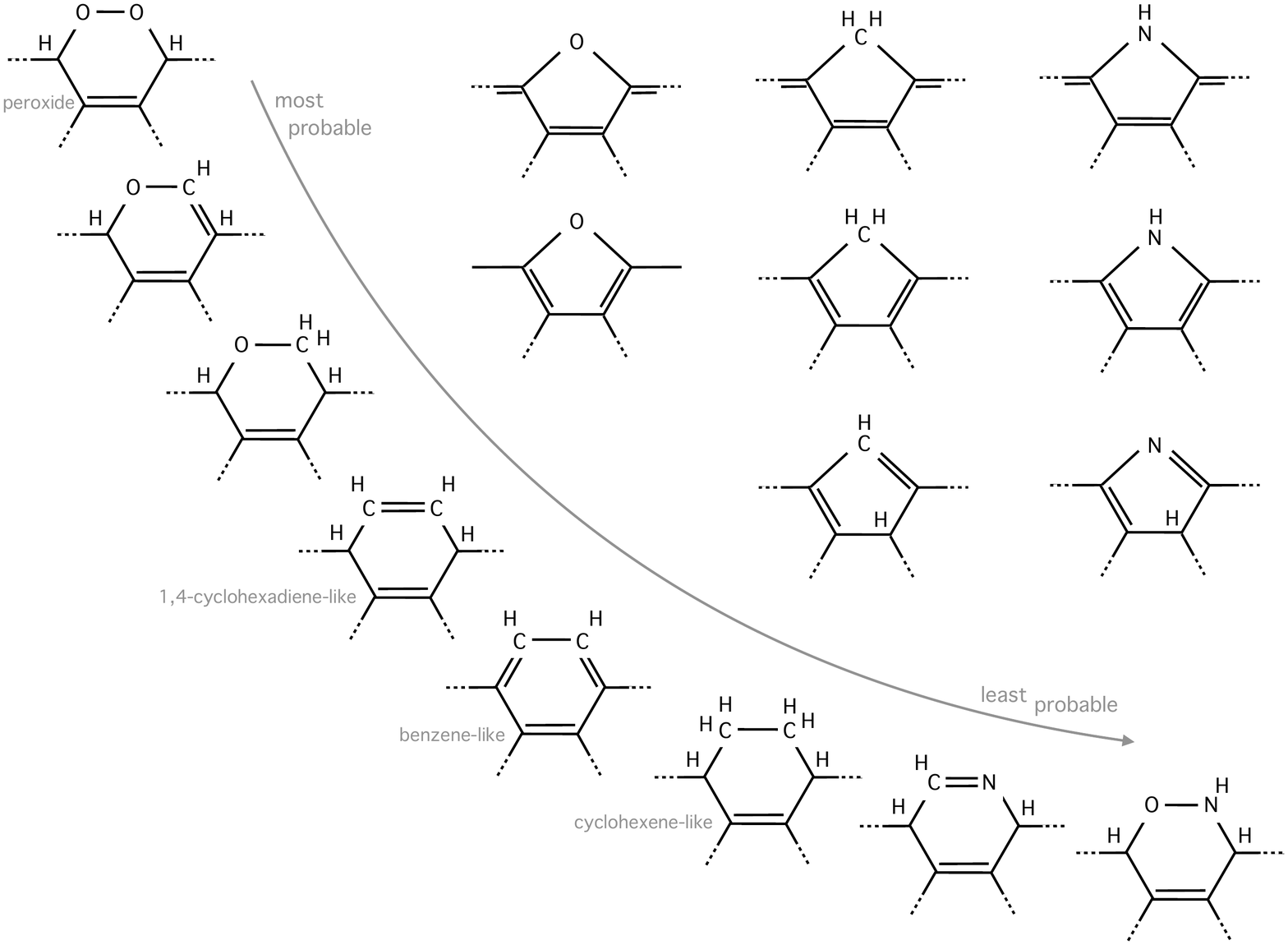}
\caption{The most probable hydrocarbon and hetero-cyclic ring structures formed via reactions with gas phase O, C and N atoms.}
\label{fig_ring_possibilities}
\end{figure}

The possible reaction pathways for gas phase carbon and hydrogen atoms with surface epoxides, accompanied by UV photolysis, are shown in \ref{fig_carbon_1}. 
In this case the most probable product species will likely be small hydrocarbon radicals ({\it e.g.}, CH$_3$, methyl), unsaturated molecules ({\it e.g.}, C$_2$H$_2$, ethyne) and small linear and cyclic hydrocarbons ({\it e.g.}, $l$-C$_3$H$_2$, propenylidene, and $c$-C$_3$H$_2$, cyclopropenylidene). 
Surface reactions can also lead to grain growth and five- and six-fold ring formation, which can form hetero-cyclic rings containing O and N atoms, as illustrated in Figs.~\ref{fig_ring_formation} and \ref{fig_ring_possibilities}. Such species could be the DIB-carrying sub-structures proposed to exist within a-C:H:X grains \citep{2013A&A...555A..39J,2014P&SS..100...26J,2016RSOS....360223J}.  
Surface epoxide species could therefore be at the heart of an extremely rich chemistry in the ISM, which needs to be explored in detail in the laboratory and theoretically. 

\begin{table}[!h]
\caption{Gas phase products arising from the UV photolysis of epoxide-activated a-C:H grains ({\it i.e.}, a-C:H:O) via  reactions with gas phase atoms (O, C and N) in low-density, atomic clouds and with polyatomic radicals and molecules (OH, CO, C$_2$, C$_2$H and CN) in denser, translucent clouds. For aliphatic (olefinic) hydrocarbon products $n [m] = 0, 1, 2$ or 3 [...] ($= 0, 1$ or 2 [...]) and for N-containing products $p = 0, 1$ or 2 ($ = 0$ or 1 for =N species). The probable abundances of the species decrease from top to bottom and also from left to right.}
\label{table_products}
\begin{tabular}{lllllll}
\hline
                & \multicolumn{6}{l}{atomic cloud products}   \\
 reactants  & \multicolumn{6}{l}{simple hydrides and diatomic oxides }   \\
\hline
 O   & OH & O$_2$ &  & &   \\
 C   & OH & CO & CH$_n$ &  &  &  \\
 N   & OH & NO & NH$_p$ &  &  &  \\
\hline
                  & \multicolumn{6}{l}{translucent cloud products}  \\
 reactants  & \multicolumn{6}{l}{simple and polyatomic hydrides and oxides}  \\
\hline
 OH              & OH & O$_2$             & HO$_2$                 &  &   \\
 CO              & OH &  CO$_2$          & HCO                      &  &  \\
 C$_2$(H)    & OH & H$_n$CCO      & CH$_n$CH$_m$  &  &   \\
 CN              & OH & H$_n$OCN      & H$_n$CNH$_p$    &  & &  \\\hline
\end{tabular}
\end{table}

\newpage
\noindent {\rhfont \Large 9. Amino acids and interstellar chemistry}\\[-1.1cm]
\section{Amino acids and interstellar chemistry}   
\label{sect_amino_acids}

The basis of almost all amino acids is the structure $^{\rm HO}_{\rm \hspace*{0.2cm}O}\hspace*{-0.05cm} \geqslant$C$-$CH(R)$-$NH$_2$, where R is other than an H atom the $-$CH(R)$-$ group is a stereo centre and the molecule exists in D- and L- forms. 
Ignoring the H atoms, the key element in the amino acid basis structure can simply be written as O$_2$CCN. 
The simplest amino acid is glycine (R = H), and examples of other simple amino acids are alanine (R = CH$_3$), valine (R = CH(CH$_3$)$_2$), leucine (R = CH$_2$CH(CH$_3$)$_2$) and isoleucine (R = CH(CH$_3$)CH$_2$CH$_3$). The R group may include five and/or sixfold aromatic rings, alcohols, thiols and additional carboxylic acid and amine functional groups. 

Amino acids can be reduced (H addition and/or O loss) to alcohols or, in the opposite sense, alcohols can be oxidised (O addition and/or H loss) to amino acids. For example, with the nomenclature ( reactant : product ), or equivalently ( addition : elimination ) for the forward reaction ($\rightarrow$) or ( elimination : addition ) for the back reaction ($\leftarrow$), these reactions can be expressed rather compactly as: \\ \\
$^{\rm HO}_{\rm \hspace*{0.2cm}O}\hspace*{-0.05cm} \geqslant$C--CH(R)--NH$_2$ \ \ $\leftarrow$ \ ( 2H : O ) \ $\rightarrow$ \ \ HO--CH$_2$--CH(R)--NH$_2$ \\ \\ 
with further oxidation (possibly as a result of photolytic dehydrogenation in the ISM) it is possible to extend this scheme to the formation of highly unsaturated species {\it i.e.}, \\ \\ 
$\leftarrow$ \ ( h$\nu$ : 4H ) \ $\rightarrow$ \ \ O=CH--CR=N--H \ \ $\leftarrow$ \ ( h$\nu$ : H, R ) \ $\rightarrow$ \ \ O=CH--C$\equiv$N. \\ \\
Comparing denatured amino acids with the interstellar OCN species, and surface epoxide chemistry, as discussed in the previous sections, it would appear that there could be a chemical structure connection between them. However, this is not to say that the observed interstellar species derive from denatured amino acids but rather the reverse. Thus, the observation of the amino acid glycine in the ISM and the numerous amino acids in primitive meteorites ({\it e.g.}, over 50 amino acids have been identified in the Murchison meteorite) indicate that they could be derived from a reduction of observed interstellar species in dense cloud cores and proto-planetary discs en route to planet formation. 

\begin{figure}[!h]
\centering\includegraphics[width=5.0in]{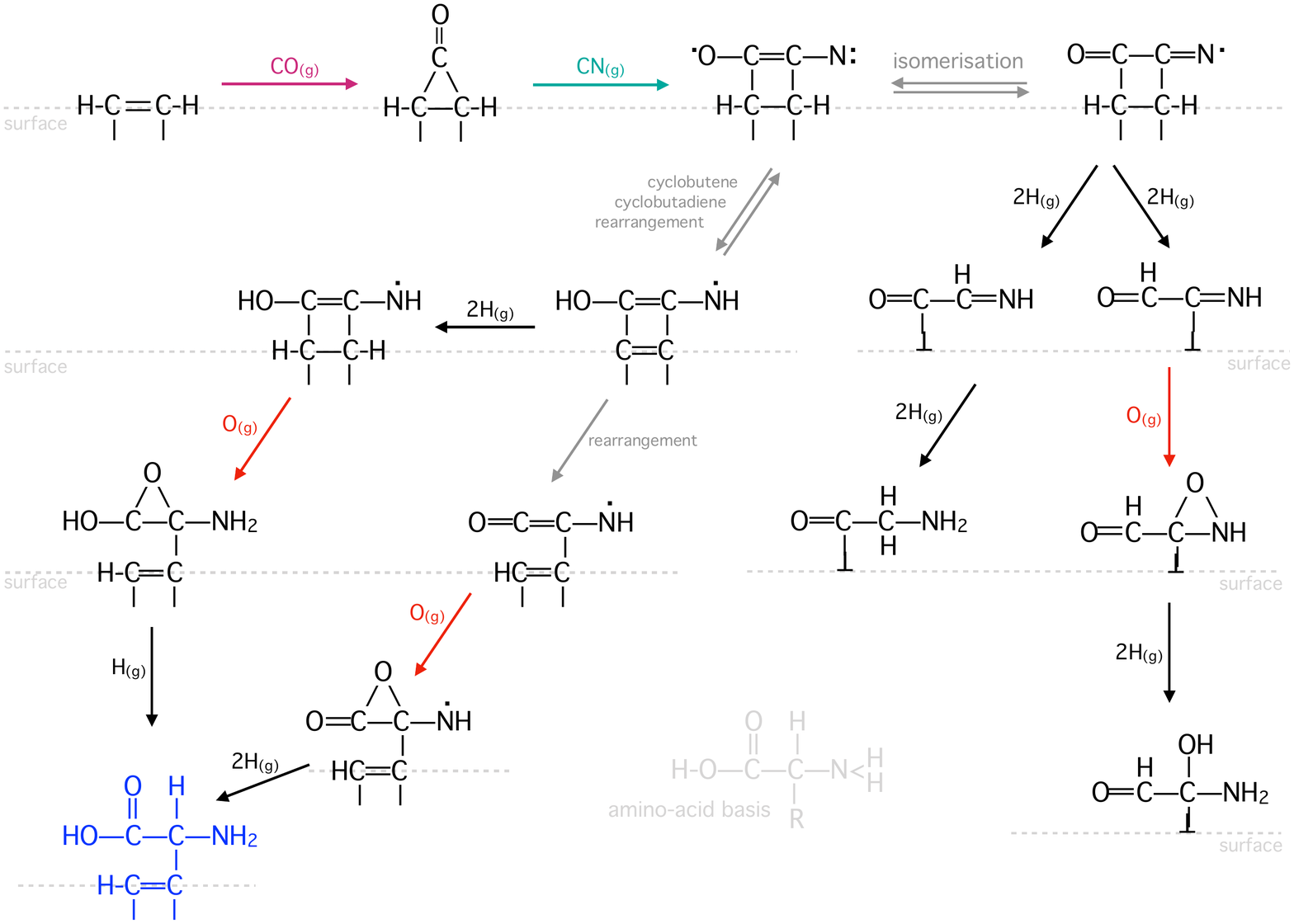}
\caption{A possible route to amino acid basis formation (lower left, blue) on a-C:H grain surfaces via reactions with gas phase CO, CN, O and H. It should be noted that the numerous branching possibilities (many more than shown) indicate that any amino acid formation on such surfaces is possible but is not likely to be very efficient. Indeed, the isomerisation of likely precursors (right hand side) leads to split channels leading to other complex organic surface species.}
\label{fig_amino_acid_formation_1}
\end{figure}

Fig.~\ref{fig_amino_acid_formation_1} illustrates a possible route to amino acid  formation on a-C:H grain surfaces in reactions with gas phase CO, CN, O and H. 
In the low density outer translucent regions of molecular clouds, where a favourable interplay between gas density and the triggering role of UV photolysis exists, the formation of amino acids and their analogues could most likely proceed through something like the following sequential pathway (see Fig.~\ref{fig_amino_acid_formation_1}). This could begin with the formation of cyclo-propanone-like surface-bonded structures via C and O atom, or CO molecule, a-C:H surface addition reactions with these abundant gas phase species. The next necessary step would be the reaction of these surface species with less abundant CN radicals through a (UV photolysis-induced) triggered ring-opening reaction to form OCCN surface-bonded structures (upper and middle part of Fig.~\ref{fig_amino_acid_formation_1}). Subsequent reactions with abundant O and H atoms (middle to lower left part of Fig.~\ref{fig_amino_acid_formation_1}) could then lead to the formation of O$_2$CCN amino acid basis structures (blue structure in the lower left of Fig.~\ref{fig_amino_acid_formation_1}). 

Thus, amino acid formation on grain surfaces is therefore theoretically possible but it is not likely to be  efficient given the numerous branching possibilities. Indeed, the isommerisation of amino acid precursors will lead to pathway-splitting and the formation of other complex organic surface species.
For example, it has been shown that high energy irradiation of amines can lead to the abiotic synthesis of amino acids, purine and pyrimidine via carboxylation \citep{Cataldo:2017kg}. Further, this same work also showed that proteinogenic amino acids and some non-proteinogenic amino acids could survive for up to 4.6 Gyr buried within comets and asteroids.

\vspace*{1.5cm}
\noindent {\rhfont \Large 10. Tying the loose ends together}\\[-1.0cm]
\section{Tying the loose ends together} 
\label{sect_loose_ends}

Linking all of the above ideas into a coherent and global view of interstellar dust might at first  seem like an insurmountable task but, once embarked upon, actually turns out to be rather straight forward. This is because the key and overriding concept is hydrocarbon chemistry (with trace amounts of O and N) and how these materials form and evolve in the ISM. The traces of O and N hetero-atoms turn out to be critically-important for determining the nature and the properties of these interesting materials. 
Here we cannot just consider dust particles, for which an exhaustive definition is not possible, because we have to include everything from simple organic molecules with only a few heavy atoms up to cometary, asteroidal and planetary-sized bodies within our global view. 

A global view of the evolutionary cycle of carbonaceous matter in space must begin somewhere, and in this case it will start (and end) with a consideration of the properties of the smallest interstellar species, {\it i.e.}, nano-particles and molecules. 
Beginning with nano-particles, which must exhibit structures significantly more complex than interstellar PAHs in order to be consistent with the various observations attributed to them. It is clear that they must be of mixed polycyclic aliphatic, olefinic and aromatic composition in order to explain the IR emission spectra and that the aromatic component may not be as prevalent as the A in PAH implies. They must be more like the mixed structures that have been described elsewhere \citep{2012A&A...542A..98J,2012ApJ...761...35M,2013A&A...558A..62J} and they are likely to make a significant contribution to the formation of molecular hydrogen and other small species in the diffuse ISM and in moderately-excited PDRs \citep{2014A&A...569A.119A,2015A&A...584A.123A,2015A&A...581A..92J,2016RSOS....360221J} and could also be the carriers of some of the diffuse interstellar bands (DIBs) \citep{2014P&SS..100...26J,2016RSOS....360223J}. Their importance for ISM physics, including gas heating via photo-electric emission, and their role in regulating the energy budget and chemistry in the low-density ISM cannot be underestimated.  Their importance and the diversity of their properties is in no small part due to their structural and chemical complexity, a complexity way beyond even the most complex PAH structures imaginable. 
This complexity carries over as the nano-particles, and all larger grains, transit to denser regions accreting complex aliphatic-rich mantles on a-C:H a material that shows the related but significantly-different IR spectral signatures typical of the dust seen along lines of sight towards the Galactic Centre (see Fig.~\ref{fig_cyclics_spectra} for a schematic view of these spectral differences) \citep{2012A&A...537A..27G,2012A&A...540A...2J,2013A&A...558A..62J,2016RSOS....360224J}. To this complexity we must now add a further level of complexity because these materials are not simply bi-atomic hydrocarbon phases but are solids that will naturally incorporated many different hetero-atoms, but principally and of most importance for ISM chemistry, DIBs and precursor biotics are O, N and S atoms \citep{2013A&A...555A..39J,2014P&SS..100...26J,2016RSOS....360224J}.  

This changing dust chemistry and the influence of dust on chemistry is obviously accompanied by accretion and a significant increase in the dust mass, and also by coagulation aided by the formation a-C:H mantles. These effects lead to many other changes in the observed properties of dust that can be interpreted through the detailed application of physically-realistic dust models, such as THEMIS. These evolutionary changes include: a net increase in the grain sizes (and the mean dust mass), initially a steepening of the FUV extinction and a loss of the UV bump followed by the loss of UV extinction yielding a grey or flat extinction from visible to shorter wavelengths, and a loss of the IR emission bands and the MIR continuum to produce a single-peaked FIR thermal dust emission SED (with a dust composition and structure-dependent shape that will be a function of the gas density and radiation field intensity). As has already been shown, it is the combination of these chemical and physical changes in the dust that are able to explain the observed properties of the dust in the outer reaches of molecular clouds \citep{2016A&A...588A..43J,2016A&A...588A..44Y,2017A&A...605A..99T}.  In essence, the sequence of carbonaceous dust chemical evolution from a-C dominated in the low density diffuse ISM to a-C:H dominated in outer molecular clouds regions and towards the Galactic Centre is accompanied by a structural evolution towards mantled grains (steeper FUV extinction, no UV bump) and then mantled-coagulated grains (grey extinction, which is flat over UV to FUV wavelengths). 

The most obvious and widely-used measure in extragalactic studies that tracks these dust evolutionary changes is the observationally-determined gas-to-dust ratio (G/D). The above discussion (see Section~5.1) showed that for a solar metallicity system the lowest possible value of G/D is of the order of $50-60$ but that the maximum value is a sensitive measure of the energy content of the system. The lowest values of G/D ($50-60$ for solar metallicity) are only possible in the densest regions of the ISM where water ice-dominated outer dust mantles can form. In the regions just exterior to the densest and coldest ISM phases, where there is no or little ice frosting on the grains, the G/D values are more typically of the order of $80-100$ and indicate regions where basically most of the condensible and more refractory elements ({\it e.g.}. O and C, and also N and S) are incorporated into dust, mostly in the form of a-C:H materials with O, N and S atoms bound within the intrinsic polycyclic ring components of these materials but also present as carboxy- ($>$C$=$O), thio- ($>$C$=$S), amine ($-$NH$_2$) and imine ($=$N) groups. Overall the accretion parameter $\xi = ( n_{\rm H} \surd T_{\rm g})^{-1}$ appears to be a reasonable indicator that dust formation is dominated by accretion in the densest ISM phases, with deviation from this tendency only seen in the most energetic or harshest of environments. 
However, the problems with the derivation of G/D in galaxies are numerous, but principle among them are that the dust mass is obviously model dependent and one has to be careful to ensure that both of the derived gas and dust masses are independent of data resolution and that they track the gas and dust masses over exactly the same regions when measured on global galactic scales. 

Within our own solar system it now appears that the G/D values, and more specifically the carbonaceous matter volume fraction typical of the dense ISM, as predicted by THEMIS dust modelling approach ({\it i.e.}, $71-74$\%) are entirely consistent with the chemical composition of comet 67P/Churyumov-Gerasimenko made by the ROSETTA mission, which found that the minimum carbonaceous material content is 75\% by volume. This might then be regarded as an {\it a posteri} confirmation of the coherency of the THEMIS global approach to dust evolutionary modelling and thus to the importance of the chemical compositional and structural evolution of  dust throughout the ISM. This connection indeed appears to be more intimate than a simple volume-counting of carbonaceous matter because, as shown in the previous paper in this series \citep{2016RSOS....360224J},  there is a fundamental link between the kinds of chemical species likely to form on chemically-active a-C:H (nano-)particle surfaces and the molecules detected in comets. This subject was again touched upon here (see sections~\ref{sect_OCN} and \ref{sect_epoxides}) when the origin and importance of OCN molecules among the detected interstellar molecules was underlined. 

Finally, and perhaps of some as yet incompletely-understood significance, is the link made in section~\ref{sect_amino_acids} between active carbonaceous a-C:H grain surface chemistry and the formation of amino acids, which again appears to link to the importance of the detected OCN species. What this work suggests is that chemical pathways to complex organic molecules probably exist on grain surfaces and that these are most likely to be active in the intermediate cloud regions, most likely the outer surfaces of molecular clouds where both accretion and UV photolysis achieve a finely-balanced dance leading to the formation of interesting and complex organics. 
Thus, and even if amino acids are not formed in quantity in space, the seed processes for their formation are likely prevalent in space with perhaps OCN species being an indicator of the importance of the possible amino acid seeding pathways. However, and while amino acids may not form directly in the outer reaches of molecular clouds, their precursor dehydrogenated (oxidised) variants can likely form in regions of active grain surface chemistry within moderately dense gas ($10^3-10^4$ H atom/cm$^{-3}$) and muted interstellar radiation fields ($A_{\rm V} \simeq 1.5-2$ \ $\equiv$ \ $A_{\rm FUV} \simeq 5$, {\it i.e.}, regions with UV photolysing fluxes reduced by orders of magnitude compared to the ambient ISM). Then, deeper into the clouds and before most of the hydrogen atoms are incorporated into molecular hydrogen, or within the outer reaches of planet-forming discs within molecular clouds, these precursor oxidised variants will be at least in part reduced through cold H atom addition/incorporation leading to H-richer species that will more closely resemble the amino acids detected within cometary and asteroidal matter. 

If the formation of OCN molecules and amino acid precursors is indeed a fundamental aspect of interstellar dust chemistry in the outer reaches of molecular clouds, where a quasi-equilibrium UV photon-driven chemistry occurs, this implies that these chemical pathways are universal and ought then to be at the basis of processes leading to the formation of life everywhere in the Universe. Thus, interstellar grain surface chemistry in intermediate regions could be the driver of a commonality in the formation of the fundamental building blocks of life, as discussed in the first paper of this series \citep{2016RSOS....360221J}. 

Thus, we end our cyclic evolutionary and global view of the dust variations observed within the ISM with a return to small organic species, which may exist as both distinct molecular entities and also as intrinsic components on the surfaces and within carbonaceous grains and the mantles on all interstellar grains. These provide not only intriguing links with interstellar chemistry but also, through hetero-atomic hydrocarbon cycles ({\it i.e.}, with incorporated O, N and S atoms), may be a key element in the elucidation of the carriers of the diffuse interstellar bands and the evolution of life itself. 

\vspace*{1.5cm}
\noindent {\rhfont \Large 11. Consequences and implications}\\[-1.1cm]
\section{Consequences and implications}   
\label{sect_implications}

In the light of the discussions of the preceding sections, it would appear that a few fundamental aspects of the nature and evolution of dust in the ISM must now be given very careful re-consideration by the community. Here a few of the more important topics for detailed re-analysis are emphasised. 

It is now perhaps opportune to undertake a re-consideration of the properties of the smallest interstellar grains, the nano-particles responsible for the IR emission bands. These bands are usually attributed to the so-called "interstellar PAHs". While PAHs in the strictest sense have long been studied within the context of astrophysics, there has yet to be any conclusive evidence that such particles make up the most abundant form of carbonaceous matter in space. Thus, and while much fascinating theoretical and experimental work has been undertaken on PAHs by the many groups working in this area, the fruits of these labours have yet to seed. The same may now be said for the burgeoning field of interstellar fullerene studies, which while interesting, miss the point that these materials are, like their PAH sisters, almost certainly a distraction from the majority of carbonaceous matter in the ISM that exists in other more complex and much more interesting forms, most likely of mixed polycyclic aliphatic, olefinic and aromatic structure. 

It is now also very clear that when we interpret the observed spectral energy distributions (SEDs) of different regions within the ISM of the Milky Way, and the different interstellar regions observed in external galaxies, that it is no longer sufficient to simply scale a "standard" interstellar radiation field and to use the same dust compositional and and structural properties everywhere. We know that the dust properties evolve with the local environment and so we must allow for this in SED modelling by including a coupled density/radiation field dependence to the dust optical properties within the model. This should, ideally, include at least different dust properties for diffuse, translucent and dense molecular clouds. In the parlance of the THEMIS model, this corresponds to the CM, CMM and AMM(AMMI) dust components or, in more detail, core/a-C mantle (CM), core/a-C mantle/a-C:H mantle (CMM), and aggregate/a-C mantle/a-C:H mantle(aggregate/a-C mantle/a-C:H mantle/ice mantle) [AMMI(AMMI)] grains. 

The measured gas-to-dust (G/D) ratios are another case in point. Firstly, it is not and probably never will be possible to determine absolute G/D values because, at least from the dust mass determination point of view, the dust emissivity used to derive a dust mass is completely model dependent. Secondly, derived gas masses must be complete ({\it i.e.}, include all of the ionised, atomic and molecular gas tracers, such as atomic H$^+$, H, H$_2$, CO, \ldots), a situation that is likewise almost impossible to achieve because the derived gas masses depend on the resolution of the observations. Thus, it is practically impossible to derive measures of both dust and gas masses over exactly the same regions. Our best current hope is therefore to consider differential measurements of G/D ratios between regions within a given galaxy and between different galaxies.   

Following from the above it should be obvious that the evolution of dust encompasses differences in the dust physical and chemical properties as it transits between regions with very different local conditions of gas density and radiation field. The chemical changes are particularly significant and would indicate a very important role for surface-driven reaction pathways to species that are not easily formed via gas phase routes. Further, these pathways and their products are not invariant and will evolve because the chemically-reactive surfaces available to the gas phase adherents (atoms, molecules and radicals) will be a function of the local conditions. In the low density ISM the surfaces will be rather olefinic/aromatic-rich a-C but in denser and mildly-extinguished regions the surfaces will be aliphatic-rich and exhibit a selectively-different surface to incident gas phase species. The full details of these reactions have yet to be fully realised and incorporated within interstellar chemical reaction networks. However, the viability of the reaction pathways proposed here have yet to be clearly demonstrated by laboratory experiments.

\vspace*{1.5cm}
\noindent {\rhfont \Large 12. Conclusions}\\[-1.1cm]
\section{Conclusions}   
\label{sect_conclusions}

Interstellar dust is an esoteric field of interest but, as has hopefully been shown above, its nature and the evolution of its chemical and structural properties with environment may have far-reaching effects. Dust studies are here discussed with a wide-ranging but shallow approach in mind with the aim of elucidating the global connections between the dust in environments encompassing the solar system, cometary, circumstellar, interstellar and extra-galactic regions. 

The major conclusions of this work can perhaps be best summarised a follows: 
\begin{enumerate}

\item A rather simple but illustrative study, using (poly)hexa-cyclic species seems to indicate that the origin and nature of the infrared absorption and emission band carriers is, as yet, far from resolved. The responsible species can seemingly be much less aromatic than has previously been considered but, nevertheless, they are likely to be structurally much more complicated than simple PAHs. Indeed, they probably consist of mixed polycyclic olefinic and aliphatics domains, which likely play a role as equally important as any polycyclic aromatic domains. 
 
\item In the ISM the existence and accretion of carbonaceous, a-C(:H), mantles onto all grains most probably precedes the coagulation of grains in the denser ISM.  Additionally, the accretion of mantles will likely aid the coagulation process by providing accommodating surfaces for grain-grain sticking. In the outer regions of molecular clouds the mantles will be hydrogen-rich a-C:H and their formation will initially shift the dust extinction to shorter wavelengths and suppress the UV bump, shifting the dust IR emission into the MIR. Subsequently, and in denser regions, the extinction will become grey and the dust emission dominated by a single-peaked SED. 
 
\item The accreted a-C:H mantles, in the outer regions of molecular clouds with moderate densities and attenuated radiation fields, will provide chemically-active substrates that can actively participate in rather rich chemical pathways. With the mediation of epoxide groups, formed by O atom reaction with surface C$=$C bonds to form epoxide-type ring structures (>C$-\hspace{-0.24 cm} ^{\rm O}$C<), grain surface chemistry could yield many interesting OCN-containing complex organic molecules, which include O--C--N, O=C--N, O=C=N and O--C$\equiv$N bonding configurations. 

\item  As carbonaceous mantles form through gas phase C atom accretion, and as they incorporate gas phase O and N atoms, there is a net decrease in the gas-to-dust ratio (G/D), from 135 to 55, and an accompanying increase in the fractional volume of the carbonaceous matter in the dust from 47 to 74\%. The G/D for dense IMS regions predicted by the THEMIS interstellar dust modelling framework is consistent with the ratios found for a wide range of galaxies and, further, the fractional volume of carbonaceous matter (74\%) in dust in dense regions is consistent with the recent results for the carbonaceous matter content (75\% by volume) found for comet 67/P.  

\item A natural consequence of the chemical reactivity and specificity of epoxylated a-C:H grain/mantle surfaces is that they could, in addition to the formation of the suite of interesting interstellar OCN species, also form OCCN or O$_2$CCN bonded configurations, which in their oxidised/reduced forms could be transformed into amino acids, {\it e.g.},  $^{\rm HO}_{\rm \hspace*{0.2cm}O}\hspace*{-0.05cm} \geqslant$C$-$CH(R)$-$NH$_2$. If the proposed diffuse/translucent interstellar medium epoxide-driven reaction pathways can be verified as a viable route to precursor or basis amino acid structure formation this would imply the existence of a universal route to their formation and the likelihood that life formed elsewhere in the Universe may also be based upon amino acids and their complex derivatives RNA and DNA. 

\end{enumerate}

Perhaps the overriding conclusion of this work is that the inherent complexity of the chemistry and physics of dust, in all its guises and in all environments where it exists and survives, has not yet been fully appreciated. Further, and because the nature of dust depends upon its environment, as complete an understanding as possible of the evolution of dust, both spatially and temporally, and an appreciation of the complex interplay between dust, gas and environment is essential. This can clearly only be brought about by a laboratory-based campaign to test and fully explore the nature of the whole complex family of carbonaceous materials and hence of likely dust analogues in space. However, the laboratory is not and never can be a perfect replicator of interstellar conditions and so there will never likely be easy or complete agreement between interstellar observations and the results of laboratory experiment. This then implies that we also need to closely integrate a theoretical roadmap into the analysis of both space and laboratory observations. One particularly striking example of where this approach is being used but has yet to reach maturity is in developing our understanding of the nature of interstellar carbonaceous nano-particles. These tricky beasts are indeed more complex than our current view of them because there is as yet no perfect agreement between their observed, measured and predicted IR spectroscopic signatures. However, this does not prevent us from using them as empirical probes of the ISM to determine such fundamentals as star formation rates and gas-to-dust ratios. Nevertheless, and without a better view of their chemistry and physics we are under-using them as key probes of ISM chemistry and of the physical conditions within the Milky Way and distant galaxies. 






\vspace*{0.5cm}
%
%
%
%

\noindent{\sf Acknowledgment}
A big thank you to all of my colleagues for all of the interesting discussions about interstellar dust over the many years.

This research was, in part, made possible through the financial support of the Agence National de la Recherche (ANR) through the programmes Cold Dust (ANR-07-BLAN-0364-01) and CIMMES (ANR-11-BS56-0029) and, currently, through the European Union's Seventh Framework Programme (FP7/2007- 2013) funding of the project DustPedia (grant agreement no. FP7-SPACE-606847).

\vspace*{1.5cm}
\noindent {\rhfont \Large References}\\[-1.1cm]


\footnotesize{
\bibliography{Ant_bibliography} 
\bibliographystyle{aa} 
}

\end{document}